\documentclass[aps,pra,twocolumn,superscriptaddress,10pt,floatfix,footinbib]{revtex4-1}

\usepackage[utf8]{inputenc}
\usepackage[TS1,T1]{fontenc}
\usepackage{microtype}
\usepackage[english]{babel}
\usepackage{amsmath}
\usepackage{amssymb}
\usepackage{graphicx}
\usepackage[breaklinks=true, colorlinks, pdfpagelabels, hidelinks]{hyperref}
\hypersetup{colorlinks,linkcolor=blue,citecolor=blue,urlcolor=blue,final}
\usepackage{cleveref}
\usepackage{dsfont}
\usepackage{fixmath}
\usepackage{tensor}
\usepackage{braket}
\usepackage[dvipsnames]{xcolor}
\usepackage{booktabs}
\usepackage{siunitx}
\usepackage{tabularx}
\usepackage{ushort}
\usepackage{scalerel}

\crefname{section}{Sec.}{Secs.}
\crefname{appendix}{Appendix}{Appendices}
\crefname{equation}{Eq.}{Eqs.}
\Crefname{equation}{Equation}{Equations}
\crefname{table}{Table}{Tables}
\crefname{figure}{Figure}{Figures}

\newcommand{\orcidicon}[1]{\href{https://orcid.org/#1}{\includegraphics[width=0.8em]{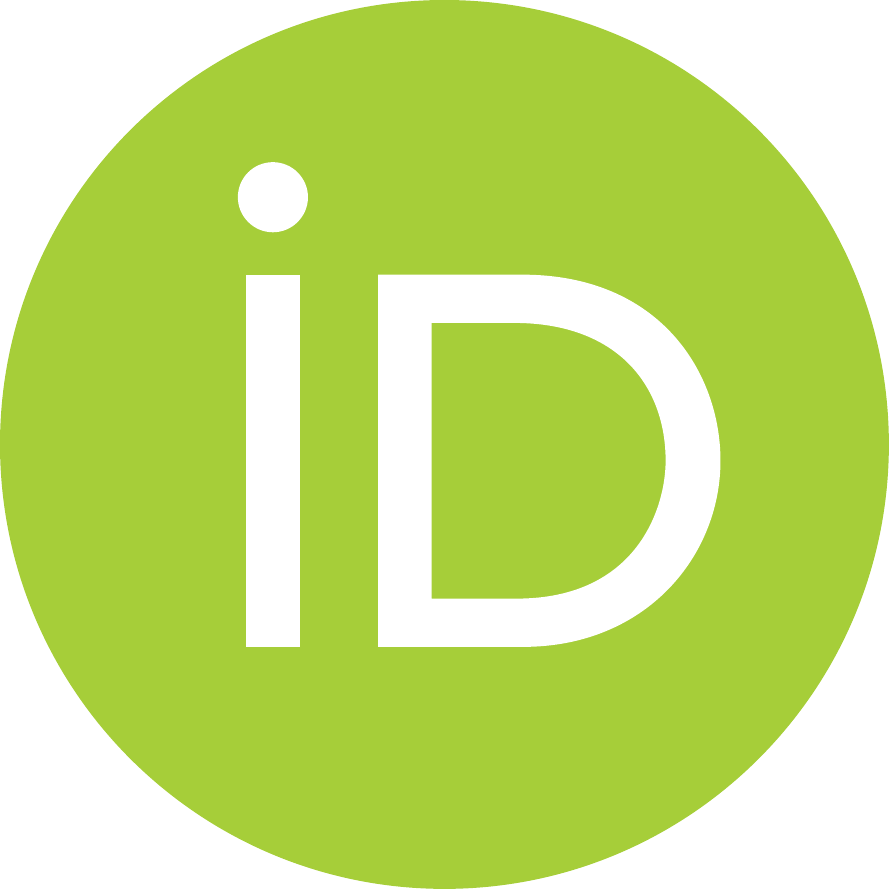}}}
\newcommand{\latin}[1]{\textit{#1}}
\newcommand{\normveccomp}{n}

\newcommand{\poscomp}{r}
\newcommand{\pos}{\vec{\poscomp}}
\newcommand{\tm}{t}
\newcommand{\post}{\pos,\tm}
\newcommand{\unitveccomp}{e}
\newcommand{\unitvec}{\vec{\unitveccomp}}
\newcommand{\xpos}{x}
\newcommand{\ypos}{y}
\newcommand{\zpos}{z}
\newcommand{\exvec}{\unitvec_\xpos}
\newcommand{\eyvec}{\unitvec_\ypos}
\newcommand{\ezvec}{\unitvec_\zpos}
\newcommand{\eAvec}{\unitvec_1}
\newcommand{\eBvec}{\unitvec_2}
\newcommand{\eCvec}{\unitvec_3}
\newcommand{\rpos}{r}
\newcommand{\polpos}{\theta}
\newcommand{\azpos}{\varphi}
\newcommand{\ervec}{\unitvec_\rpos}

\newcommand{\besselj}[1]{j_{#1}}
\newcommand{\legendreP}[2]{P_{#1}^{#2}}
\newcommand{\sphericalY}[2]{Y_{#1}^{#2}}
\newcommand{\VSHYcomp}{Y}
\newcommand{\VSHPsicomp}{\Psi}
\newcommand{\VSHPhicomp}{\Phi}
\newcommand{\VSHY}[2]{\vec{\VSHYcomp}_{#1}^{#2}}
\newcommand{\VSHPsi}[2]{\vec{\VSHPsicomp}_{#1}^{#2}}
\newcommand{\VSHPhi}[2]{\vec{\VSHPhicomp}_{#1}^{#2}}
\newcommand{\generalizedHypergeometricF}[2]{\tensor[_{#1}]{F}{_{#2}}}
\DeclareMathOperator{\real}{Re}

\renewcommand{\vec}[1]{\mathbold{#1}}
\newcommand{\zerovec}{\boldsymbol{0}}
\newcommand{\tens}[1]{\vec{#1}}
\newcommand{\zerotens}{\boldsymbol{0}}
\newcommand{\transposesymbol}{t}
\newcommand{\transp}[1]{{#1}^\transposesymbol}
\newcommand{\ccsymbol}{*}
\newcommand{\cconj}[1]{{#1}^\ccsymbol}
\newcommand{\hcsymbol}{\dagger}
\newcommand{\hconj}[1]{{#1}^\hcsymbol}
\newcommand{\op}[1]{\hat{#1}}
\newcommand{\kronecker}{\delta}
\newcommand{\dd}{d}
\newcommand{\im}{i}

\newcommand{\hc}{\text{H.c.}}
\newcommand{\cc}{\text{c.c.}}
\newcommand{\C}{\mathds{C}}
\newcommand{\R}{\mathds{R}}
\newcommand{\Z}{\mathds{Z}}
\newcommand{\N}{\mathds{N}}
\newcommand{\id}{\mathds{1}}
\DeclareMathOperator{\diagonal}{diag}
\newcommand{\grad}{\nabla}
\newcommand{\nablacomp}{\triangledown}
\newcommand{\del}{\partial}

\newcommand{\fpd}[2]{\frac{\partial {#1}}{\partial {#2}}}

\newcommand{\ddelt}{\dot}
\newcommand{\ddeltt}{\ddot}
\newcommand{\pare}[1]{\left( {#1} \right)}
\newcommand{\spare}[1]{\left[ {#1} \right]}
\newcommand{\cpare}[1]{\left\{ {#1} \right\}}
\newcommand{\trans}{t}
\newcommand{\longitud}{l}
\newcommand{\thermal}{\mathrm{th}}

\newcommand{\mass}{m}

\newcommand{\potinternal}{V}
\newcommand{\elasenergydens}{\mathcal{V}}
\newcommand{\elasenergyop}{\op{\potinternal}}
\newcommand{\potexternal}{V_\text{ext}}
\newcommand{\Lagrangian}{L}
\newcommand{\Hamildens}{\mathcal{H}}
\newcommand{\Hamiltonian}{H}
\newcommand{\rad}{R}
\newcommand{\halfAxis}{a}
\newcommand{\halfAxisA}{\halfAxis_\xpos}
\newcommand{\halfAxisB}{\halfAxis_\ypos}
\newcommand{\halfAxisC}{\halfAxis_\zpos}
\newcommand{\geometricFactor}{L}
\newcommand{\geometricFactorA}{\geometricFactor_\xpos}
\newcommand{\geometricFactorB}{\geometricFactor_\ypos}
\newcommand{\geometricFactorC}{\geometricFactor_\zpos}
\newcommand{\body}{B}
\newcommand{\dens}{\rho}
\newcommand{\lamemu}{\mu}
\newcommand{\lamelambda}{\lambda}
\newcommand{\Poissonnu}{\nu}
\newcommand{\ufieldcomp}{u}
\newcommand{\ufield}{\vec{\ufieldcomp}}
\newcommand{\elastenscomp}{C}
\newcommand{\elastens}{\tens{\elastenscomp}}
\newcommand{\straintenscomp}{S}
\newcommand{\straintens}{\tens{\straintenscomp}}
\newcommand{\averageStrain}{\bar{\straintens}}
\newcommand{\averageStrainComp}{\bar{\straintenscomp}}
\newcommand{\stresstenscomp}{T}
\newcommand{\stresstens}{\tens{\stresstenscomp}}
\newcommand{\pifieldcomp}{\pi}
\newcommand{\pifield}{\vec{\pifieldcomp}}
\newcommand{\Dphon}{\mathcal{D}}
\newcommand{\DphonInt}{\mathcal{D}'}
\newcommand{\wmodecomp}{w}
\newcommand{\wmode}{\vec{\wmodecomp}}
\newcommand{\wmodercomp}{\mathcal{W}}
\newcommand{\wmoder}{\vec{\wmodercomp}}
\newcommand{\strainmodecomp}{s}
\newcommand{\strainmode}{\vec{\strainmodecomp}}
\newcommand{\averageStrainModeComp}{\bar{\strainmodecomp}}
\newcommand{\averageStrainMode}{\bar{\strainmode}}

\newcommand{\stressmodecomp}{t}
\newcommand{\stressmode}{\vec{\stressmodecomp}}
\newcommand{\phonindex}{\gamma}
\newcommand{\phonindexcc}{\tilde{\phonindex}}
\newcommand{\phonindexb}{{\phonindex'}}

\newcommand{\phonfreq}{\omega}
\newcommand{\phonnormvar}{a}

\newcommand{\ufieldop}{\op{\ufield}}
\newcommand{\phonaop}{\op{a}}
\newcommand{\phonaopInt}{\op{b}}
\newcommand{\ufieldmodedens}{\mathcal{U}}
\newcommand{\Hamilop}{\op{\Hamiltonian}}
\newcommand{\HamilopBare}{\Hamilop_0}
\newcommand{\HamilopBareDisplaced}{\displaced{\Hamilop}_0}
\newcommand{\HamilopIntLinear}{\Hamilop_1}
\newcommand{\HamilopIntQuadratic}{\Hamilop_2}
\newcommand{\HamilopIntQuadraticDisplaced}{\displaced{\Hamilop}_2}
\newcommand{\clong}{c_\longitud}
\newcommand{\ctrans}{c_\trans}
\newcommand{\phona}{a}
\newcommand{\phonb}{b}
\newcommand{\phonl}{l}
\newcommand{\phonlb}{{l'}}
\newcommand{\phonm}{m}
\newcommand{\phonmb}{{m'}}
\newcommand{\rposR}{x}
\newcommand{\phonaR}{\alpha}
\newcommand{\phonbR}{\beta}
\newcommand{\phonfam}{f}
\newcommand{\phonfamb}{\phonfam'}
\newcommand{\Tmode}{T}
\newcommand{\Smode}{S}
\newcommand{\phonn}{{n}}
\newcommand{\phonnb}{{\phonn'}}
\newcommand{\inertialtenscomp}{I}
\newcommand{\inertialtens}{\tens{\inertialtenscomp}}

\newcommand{\poslab}{\vec{r}'}
\newcommand{\velocitylab}{\ddelt{\vec{r}}'}
\newcommand{\poscomovcomp}{r}
\newcommand{\poscomov}{\vec{\poscomovcomp}}
\newcommand{\CMpos}{\vec{r}_\text{cm}}
\newcommand{\CMvelocity}{\ddelt{\vec{r}}_\text{cm}}
\newcommand{\bodymass}{M}
\newcommand{\poscorotcomp}{r}
\newcommand{\poscorot}{\vec{\poscorotcomp}}

\newcommand{\rotmatrixcomp}{D}
\newcommand{\rotmatrix}{\tens{\rotmatrixcomp}}
\newcommand{\eulerangles}{\vec{\Omega}}
\newcommand{\eulerA}{\alpha}
\newcommand{\eulerB}{\beta}
\newcommand{\eulerC}{\gamma}
\newcommand{\eEulerA}{\unitvec_{\eulerA}}
\newcommand{\eEulerB}{\unitvec_{\eulerB}}
\newcommand{\eEulerC}{\unitvec_{\eulerC}}
\newcommand{\posequicomp}{R}
\newcommand{\posequi}{\vec{\posequicomp}}

\newcommand{\rotfreq}{\omega}
\newcommand{\rotfreqvec}{\vec{\rotfreq}}
\newcommand{\rotfreqcrit}{\rotfreq_c}
\newcommand{\rotfreqnl}{\rotfreq_\text{nl}}
\newcommand{\bscoupling}{k}
\newcommand{\sqcoupling}{g}
\newcommand{\linshift}{e}
\newcommand{\Hamilmatrixcomp}{M}
\newcommand{\Hamilmatrix}{\tens{\Hamilmatrixcomp}}
\newcommand{\HamilmatrixDiagonal}{\tens{M}_0}
\newcommand{\HamilmatrixHybridization}{\tens{M}_2}
\newcommand{\aopveccomp}{\op{\Psi}}
\newcommand{\aopvec}{\op{\vec{\Psi}}}
\newcommand{\relHybridizationCritical}{\chi_c}
\newcommand{\frequencyMatrixComp}{W}
\newcommand{\frequencyMatrix}{\tens{\frequencyMatrixComp}}
\newcommand{\bscouplingMatrix}{\tens{K}}
\newcommand{\sqcouplingMatrix}{\tens{G}}
\newcommand{\displop}{\op{D}}
\newcommand{\phondispl}{d}
\newcommand{\displaced}[1]{\ushortw{#1}}
\newcommand{\phonfreqInt}{\Omega}
\newcommand{\phonindexInt}{\xi}
\newcommand{\phonindexIntb}{{\phonindexInt'}}

\newcommand{\wmodeInt}{\vec{v}}

\newcommand{\dynamicalMatrix}{\tens{D}}

\newcommand{\JMatrix}{\tens{J}}
\newcommand{\trafomatrixcomp}{U}
\newcommand{\trafomatrix}{\tens{\trafomatrixcomp}}
\newcommand{\HamilmatrixInt}{\tens{N}}
\newcommand{\phonmodeoverlap}{\mathcal{A}}

\newcommand{\permitt}{\epsilon}
\newcommand{\vacpermitt}{\permitt_0}
\newcommand{\relpermitt}{\permitt_r}
\newcommand{\impermitt}{\eta}
\newcommand{\impermitttenscomp}{\impermitt}

\newcommand{\relpermitttensbcomp}{\permitt}
\newcommand{\relpermitttensb}{{\tens{\permitt}}}

\newcommand{\polarizab}{\alpha}
\newcommand{\polarizabtenscomp}{\polarizab}
\newcommand{\polarizabtens}{\tens{\polarizabtenscomp}}
\newcommand{\massindex}{\tau}
\newcommand{\massindexb}{{\massindex'}}
\newcommand{\LeviCivita}{\epsilon}
\newcommand{\pockelstenscomp}{P}
\newcommand{\photela}{\pockelstenscomp_1}
\newcommand{\photelb}{\pockelstenscomp_2}
\newcommand{\pockelstens}{\tens{\pockelstenscomp}}
\newcommand{\volumeEllipsoid}{V_\text{e}}
\newcommand{\labframe}{R_L}
\newcommand{\bodyframe}{R_B}

\DeclareMathOperator{\tr}{Tr}
\newcommand{\anharmonicA}{A}
\newcommand{\anharmonicB}{B}
\newcommand{\anharmonicC}{C}
\newcommand{\anharmonicAdimless}{\tilde{A}}
\newcommand{\anharmonicBdimless}{\tilde{B}}
\newcommand{\anharmonicCdimless}{\tilde{C}}
\newcommand{\order}{\mathcal{O}}
\newcommand{\aharmcorrection}{h}
\DeclareMathOperator{\atan}{arctan}
\renewcommand{\max}{\text{max}}
\newcommand{\tensorinvar}{\mathcal{I}}

\begin{document}
\selectlanguage{english}

\title{Acoustic and Optical Properties of a Fast-spinning Dielectric Nanoparticle}

\author{Daniel Hümmer\,\orcidicon{0000-0002-0228-2887}}
\affiliation{Institute for Quantum Optics and Quantum Information of the Austrian Academy of Sciences, 6020 Innsbruck, Austria}
\affiliation{Institute for Theoretical Physics, University of Innsbruck, 6020 Innsbruck, Austria}
\author{René Lampert}
\affiliation{Institute for Quantum Optics and Quantum Information of the Austrian Academy of Sciences, 6020 Innsbruck, Austria}
\affiliation{Institute for Theoretical Physics, University of Innsbruck, 6020 Innsbruck, Austria}
\author{Katja Kustura}
\affiliation{Institute for Quantum Optics and Quantum Information of the Austrian Academy of Sciences, 6020 Innsbruck, Austria}
\affiliation{Institute for Theoretical Physics, University of Innsbruck, 6020 Innsbruck, Austria}
\author{Patrick Maurer\,\orcidicon{0000-0002-0423-8898}}
\affiliation{Institute for Quantum Optics and Quantum Information of the Austrian Academy of Sciences, 6020 Innsbruck, Austria}
\affiliation{Institute for Theoretical Physics, University of Innsbruck, 6020 Innsbruck, Austria}
\author{Carlos Gonzalez-Ballestero\,\orcidicon{0000-0002-7639-0856}}
\affiliation{Institute for Quantum Optics and Quantum Information of the Austrian Academy of Sciences, 6020 Innsbruck, Austria}
\affiliation{Institute for Theoretical Physics, University of Innsbruck, 6020 Innsbruck, Austria}
\author{Oriol Romero-Isart\,\orcidicon{0000-0003-4006-3391}}
\affiliation{Institute for Quantum Optics and Quantum Information of the Austrian Academy of Sciences, 6020 Innsbruck, Austria}
\affiliation{Institute for Theoretical Physics, University of Innsbruck, 6020 Innsbruck, Austria}
\date{\today}

\begin{abstract}

Nanoparticles levitated in vacuum can be set to spin at ultimate frequencies, limited only by the tensile strength of the material. At such high frequencies, drastic changes to the dynamics of solid-state quantum excitations are to be expected. Here, we theoretically describe the interaction between acoustic phonons and the rotation of a nanoparticle around its own axis, and model how the acoustic and optical properties of the nanoparticle change when it rotates at a fixed frequency. As an example, we analytically predict the scaling of the shape, the acoustic eigenmode spectrum, the permittivity, and the polarizability of a spinning dielectric nanosphere. We find that the changes to these properties at frequencies of a few gigahertz achieved in current experiments should be measurable with presents technology. Our work aims at exploring solid-state quantum excitations in mesoscopic matter under extreme rotation, a regime that is now becoming accessible with the advent of precision control over highly isolated levitated nanoparticles.

\end{abstract}

\maketitle

\section{Introduction}

Levitated nanoparticles in high vacuum~\cite{romero-isart_toward_2010,chang_cavity_2010,romero-isart_optically_2011}  have attracted interest due to their excellent isolation from the environment paired with the ability to optically detect their position with high precision. Recent advances have rendered it possible to control the translational motion of such nanoparticles at the level of single quanta~\cite{delic_cooling_2020,tebbenjohanns_motional_2020, tebbenjohanns_cold_2019,gonzalez-ballestero_theory_2019,meyer_resolved-sideband_2019,windey_cavity-based_2019,delic_cavity_2019}. At the same time, there is a growing level of control over their rotational degrees of freedom~\cite{arita_laser-induced_2013,hoang_torsional_2016,kuhn_full_2017,nagornykh_optical_2017,rahman_laser_2017,monteiro_optical_2018,reimann_ghz_2018,ahn_optically_2018}. More recently, attention has been focused on the solid-state quantum excitations of nanoparticles~\cite{rusconi_quantum_2017,rubio_lopez_internal_2018,gonzalez-ballestero_quantum_2020}. The latter is motivated by the fact that well-isolated nanoparticles can be used to study solids at the mesoscopic scale, where their quantum excitations can be highly discretized and long lived~\cite{maccabe_phononic_2019}, and their dynamics may consequently be radically different from bulk solids and non-isolated mesoscopic systems. Internal solid-state quantum excitations (e.g.,\ acoustic phonons, magnons, or plasmons) should not be confused with the internal degrees of freedom of nanoparticles embedded with quantum emitters~\cite{kuhlicke_nitrogen_2014,hoang_electron_2016,juan_near-field_2016,rahman_laser_2017,delord_ramsey_2018,conangla_motion_2018}.

In this context, a key advantage of levitated nanoparticles is the possibility to spin them at the highest frequencies, limited only by the cohesion of the material~\cite{ahn_optically_2018,reimann_ghz_2018,monteiro_optical_2018,ahn_ultrasensitive_2020}. This situation offers the unique opportunity to study the mesoscopic internal physics of a nanoparticle under extreme rotation. In this paper, we start to address this research direction in the simplest case of a nanosphere and  acoustic vibrations (phonons). To this end, we model the coupled dynamics of rotational and vibrational degrees of freedom of a nanoparticle from first principles. We then analytically study how the linear acoustic and optical properties of the nanoparticle are modified when it is spinning at a fixed frequency. We find that for dielectric nanospheres rotating at gigahertz (\si{\giga\hertz}) frequencies as achieved in current experiments~\cite{reimann_ghz_2018,ahn_ultrasensitive_2020}, rotation-induced changes should be measurable, for instance in the polarizability of the nanosphere and the phonon frequency spectrum. Moreover, we find that nonlinear elastic effects could already be detected at such frequencies. We remark that related studies and experiments have been performed for levitated droplets of classical~\cite{wang_bifurcation_1994,hill_nonaxisymmetric_2008,baldwin_artificial_2015} and superfluid liquid~\cite{childress_cavity_2017,bernando_shapes_2017,langbehn_three-dimensional_2018,aiello_perturbation_2019} as well as with graphene nanoplatelets~\cite{nagornykh_optical_2017}.

This paper is structured as follows: In \cref{sec: theory}, we develop a general theory to describe the coupling of the rotational and vibrational degrees of freedom of a linear elastic rotor of arbitrary shape. In particular, we obtain the Hamiltonian governing the dynamics of the phonon field of a body spinning at a fixed frequency. We then use this model to infer how the shape, the acoustic eigenmode spectrum, and the permittivity are modified under rotation. For the particular case of a spherical particle, we also describe the change of the electric polarizability. In \cref{sec: case study}, we study the dependence of these properties on the rotation frequency for the particular case of a levitated dielectric nanosphere, using parameters corresponding to the experiment reported in Ref.~\cite{reimann_ghz_2018}. We conclude and provide an outlook for further research directions in \cref{sec: conclusions}. \Cref{sec: elastodynamics appendix} contains a review of the core concepts of elastodynamics which we use to model acoustic excitations. In \cref{sec: linear elastic rotor appendix}, we extend this theory to account for the spinning of the entire body and sketch how its general Lagrangian can be justified. In \cref{sec: sphere appendix}, we revise the phonon eigenmode structure of a resting sphere and provide details on our model for the particular case of a spinning spherical nanoparticle. In \cref{sec: nonlinearity appendix}, we describe at which rotation frequencies nonlinear elastic effects start to appear.

\section{Theoretical Modeling}
\label{sec: theory}

We model the spinning nanoparticle as a linear elastic rotor the vibrations of which can be described using linear elasticity theory. A brief introduction to linear elastodynamics is given in \cref{sec: elastodynamics appendix}. In \cref{sec: hamiltonian}, we construct the Hamiltonian of a nanoparticle spinning at a given frequency starting from the general Lagrangian of a linear elastic rotor of arbitrary shape. In \cref{sec: acoustic properties}, we discuss the influence of rotation on the shape and acoustic properties of the spinning nanoparticle by analyzing this Hamiltonian. In \cref{sec: optical properties}, we approximate how rotation modifies the optical properties of the nanoparticle due to changes in its shape.

\subsection{Spinning Nanoparticle}
\label{sec: hamiltonian}

A linear elastic rotor has translational, rotational, and vibrational degrees of freedom, described by its center of mass position $\CMpos(\tm)$, Euler angles $\eulerangles(\tm)$, and displacement field $\ufield(\post)$, respectively. In this work, we focus on nanoparticles levitated in a harmonic potential. Rotational and vibrational degrees of freedom then decouple from the center of mass and their joint evolution can be described by the Lagrangian
\begin{equation}\label{eqn: lagrangian elastic rotor}
      \Lagrangian = \int_\body \spare{ \frac{\dens}{2}  \ddelt\ufield^2 - \frac{1}{2} \straintenscomp^{ij} \elastenscomp^{ijkl} \straintenscomp^{kl}} \dd \pos + \frac{1}{2} \rotfreq^{i} \inertialtenscomp^{ij}[\ufield] \rotfreq^j
\end{equation}
which can be justified from first principles; see \cref{sec: linear elastic rotor appendix}. The first term is the standard Lagrangian of linear elastodynamics~\cite{achenbach_wave_1973,eringen_elastodynamics_1975}. The displacement field $\ufield(\post)$ indicates the time-dependent displacement of an infinitesimal volume element of the elastic body $\body$ from its equilibrium position $\pos$, relative to a comoving and corotating body frame. As detailed in \cref{sec: linear elastic rotor appendix}, comoving implies that this reference frame has the same linear velocity as the nanoparticle and corotating means it has the same angular velocity. Further, $\dens(\pos)$ is the mass density of the nanoparticle, $\elastenscomp^{ijkl}$ are the components of the elasticity tensor $\elastens(\pos)$ encoding its elastic properties, and $\straintenscomp^{ij} \equiv \pare{\partial_i \ufieldcomp^j + \partial_j \ufieldcomp^i}/2$ are the components of the strain tensor $\straintens(\post)$ that describes how the nanoparticle is deformed under a displacement field $\ufield(\post)$. Here, the indices $i,j,k$, and $l$ label the Cartesian components relative to the orthonormal basis $\{\eAvec,\eBvec,\eCvec\}$ spanning the body frame and we use the convention that repeated indices are implicitly summed over. The second term in \cref{eqn: lagrangian elastic rotor} describes the kinetic energy due to the rotation of the body around its own axis. The angular velocity vector $\rotfreqvec(\tm)$ in general varies in time via its dependence on the Euler angles $\eulerangles(\tm)$; see \cref{eqn: angular velocity body frame in lab frame}. As vibrations redistribute the mass of the rotor, the inertial tensor $\inertialtens$ is a functional of the displacement field
\begin{multline}
    \inertialtenscomp^{ij}[\ufield] \equiv \int_\body \dens \big[ (\poscomp^k + \ufieldcomp^k) (\poscomp^k + \ufieldcomp^k)\kronecker^{ij} \\
    - (\poscomp^i + \ufieldcomp^i)(\poscomp^j + \ufieldcomp^j) \big] \dd \pos
\end{multline}
where $\kronecker^{ij}$ is the Kronecker symbol and $\poscomp^i = \pos \cdot \unitvec_i$ are the components of the position vector relative to the body frame. The inertial tensor is symmetric, $\inertialtenscomp^{ij} = \inertialtenscomp^{ji}$.

The dependence of the inertial tensor on the displacement field engenders a coupling between the rotation and vibrations (phonons) of the nanoparticle. Accordingly, the equation of motion for the displacement field
\begin{equation}\label{eqn: eom displacement field}
  \dens \ddeltt\ufield(\post) = \Dphon \ufield(\post) + \dens \rotfreq^2 \pos_\perp + \dens \rotfreq^2 \ufield_\perp(\post)
\end{equation}
involves the angular velocity $\rotfreq(\tm) \equiv |\rotfreqvec(\tm)|$ of the nanoparticle, where we define the differential operator $\Dphon$ acting on the displacement field as \mbox{$\spare{\Dphon\ufield }^i = \partial_j \elastenscomp^{ijkl} \,\partial_k \ufieldcomp_l$}. The vectors $\pos_\perp$ and \mbox{$\ufield_\perp(\post)$} are the projections of the position and displacement vectors onto the plane orthogonal to the rotation axis along $\rotfreqvec$. \Cref{eqn: eom displacement field} states that the displacement field is accelerated by different force densities: The first term on the right-hand side describes the elastic restoring force opposing the displacement $\ufield(\post)$ which deforms the nanoparticle. The second and third terms correspond to the centrifugal forces that act at each point $\pos + \ufield(\post)$ of the deformed nanoparticle. There is a second set of equations of motion which describe the evolution of the Euler angles; see \cref{eqn: eom displacement field and euler angles}. Both sets of dynamical equations are coupled such that vibrations lead to a dynamical modulation of the rotation and \latin{vice versa}.

In this work, we focus on the impact that fast spinning has on the displacement field. To this end, we assume that fluctuations of the rotation frequency and axis can be neglected. This assumption is supported by recent reports on high frequency stability already for rotations in the megahertz regime~\cite{van_der_laan_optically_2020}. The frequency $\rotfreqvec$ is then a constant parameter and the displacement field $\ufield(\post)$ is the only remaining degree of freedom, governed by the dynamical equation \cref{eqn: eom displacement field}. We therefore base our discussion on the Hamiltonian corresponding to the Lagrangian \cref{eqn: lagrangian elastic rotor} in the case of a fixed rotation frequency; see \cref{eqn: Hamilton functional fixed rotation}. We express the Hamiltonian in terms of the eigenmodes $\wmode_\phonindex(\pos)$ of a nonrotating nanoparticle. These eigenmodes are eigenvectors of the differential operator $\Dphon$
\begin{equation}\label{eqn: phonon eigenmode equation}
  \Dphon \,\wmode_\phonindex(\pos) = -\dens \phonfreq^2_\phonindex\, \wmode_\phonindex(\pos)
\end{equation}
where $\phonfreq_\phonindex$ are the real-valued eigenfrequencies and $\phonindex$ is a multi-index suitable for labeling all eigenmodes.

Hereafter, we formulate the theory using quantum mechanics for the sake of generality; however, to the extent of results obtained in this paper, the analysis can be performed in an entirely analogous manner based on the classical Hamiltonian \cref{eqn: Hamilton functional fixed rotation} \footnote{In order to obtain a formulation of our analysis based on the classical theory instead of its quantized counterpart, it is sufficient to change to the Heisenberg picture and replace all ladder operators $\phonaop_\phonindex(\tm)$ with complex valued normal variables $\phonnormvar_\phonindex(\tm)$.}. After standard canonical quantization based on the eigenmodes $\wmode_\phonindex(\pos)$ (see \Cref{sec: elastodynamics appendix}), the resulting displacement field operator can be expanded in terms of the eigenmodes as
\begin{equation}\label{eqn: displacement field operator}
  \ufieldop(\pos) = \sum_\phonindex \ufieldmodedens_\phonindex \spare{ \phonaop_\phonindex \,\wmode_\phonindex(\pos) + \hc }.
\end{equation}
Here, $\ufieldmodedens_\phonindex \equiv \sqrt{ \hbar/ (2\dens \phonfreq_\phonindex)}$ is the mode density, $\phonaop_\phonindex$ are the ladder operators of the phonon field (corresponding to eigenmodes of the nonrotating nanoparticle), and $\hc$ indicates the Hermitian conjugate. The ladder operators satisfy the canonical commutation relations $[\phonaop_\phonindex,\hconj\phonaop_\phonindexb]=\kronecker_{\phonindex\phonindexb}$ subject to the proper normalization of the eigenmodes; see \cref{eqn: phonon orthonormality condition}.
The quantum Hamiltonian of an elastic body spinning at a fixed frequency then takes the form
\begin{equation}\label{eqn: fixed rotation total Hamiltonian}
  \Hamilop =\HamilopBare + \HamilopIntLinear + \HamilopIntQuadratic.
\end{equation}
The first term is the Hamiltonian of the freely evolving phonon field
\begin{equation}\label{eqn: fixed rotation total Hamiltonian contributions 1}
 \HamilopBare  \equiv \hbar \sum_{\phonindex} \phonfreq_\phonindex \hconj\phonaop_\phonindex \phonaop_\phonindex.
\end{equation}
The second and third term describe the additional centrifugal forces:
\begin{equation}\label{eqn: fixed rotation total Hamiltonian contributions 2}
  \begin{split}
  \HamilopIntLinear & \equiv  \hbar \sum_{\phonindex} \spare{ \linshift_\phonindex \phonaop_\phonindex + \hc}, \\
  \HamilopIntQuadratic & \equiv \hbar \sum_{\phonindex\phonindexb}  \bscoupling_{\phonindex\phonindexb} \hconj\phonaop_\phonindex \phonaop_\phonindexb + \frac{\hbar}{2} \sum_{\phonindex \phonindexb} \spare{ \sqcoupling_{\phonindex\phonindexb} \phonaop_\phonindex \phonaop_\phonindexb + \hc }.
  \end{split}
\end{equation}
Here, $\linshift_\phonindex$ are linear shifts that quantify the static centrifugal force $\dens \rotfreq^2 \pos_\perp$, while the beam-splitter coupling constants $\bscoupling_{\phonindex\phonindexb}$ and two-mode squeezing coupling constants $\sqcoupling_{\phonindex\phonindexb}$ quantify the dynamical centrifugal force $\dens \rotfreq^2 \ufield_\perp(\post)$. We assume without loss of generality that $\ezvec \parallel \eCvec \parallel \rotfreqvec$ where $\ezvec$ marks the $\zpos$ direction of the laboratory frame. In this case, the constants appearing in the Hamiltonian are
\begin{equation}\label{eqn: coupling constants general}
  \begin{split}
    \linshift_\phonindex &= -\frac{\dens \ufieldmodedens_\phonindex \rotfreq^2 }{\hbar} \int_\body \spare{\wmode_\phonindex \cdot \pos - \wmodecomp_\phonindex^{3} \poscomp^3} \dd \pos,\\
    \sqcoupling_{\phonindex\phonindexb} &= -\frac{\dens \ufieldmodedens_\phonindex \ufieldmodedens_\phonindexb \rotfreq^2 }{\hbar} \int_\body \spare{\wmode_\phonindex \cdot \wmode_\phonindexb - \wmodecomp_\phonindex^{3}\wmodecomp_\phonindexb^{3}} \dd \pos ,\\
    \bscoupling_{\phonindex\phonindexb} &= -\frac{\dens \ufieldmodedens_\phonindex \ufieldmodedens_\phonindexb \rotfreq^2 }{\hbar} \int_\body \spare{\cconj\wmode_\phonindex \cdot \wmode_\phonindexb - \wmodecomp_\phonindex^{3\,\ccsymbol}\wmodecomp_\phonindexb^{3}} \dd \pos
  \end{split}
\end{equation}
where $\poscomp^3 = \pos \cdot \eCvec$ and $\wmodecomp_\phonindex^{3} =\wmode_\phonindex \cdot \eCvec$. Note that the form of the Hamiltonian and the expressions for the coupling constants are obtained without needing to specify the geometry of the particle.

It is possible to obtain explicit expressions for the constants \cref{eqn: coupling constants general} in the particular case of a homogeneous and isotropic sphere. The phononic eigenmode structure of a nonrotating and freely vibrating sphere is well known; see \cref{sec: sphere appendix} for a summary. A sphere supports two distinct families $\phonfam$ of modes: torsional modes ($\phonfam = \Tmode$) and spheroidal modes ($\phonfam = \Smode$). Each  eigenmode can be labeled with a mode index $\phonindex = (\phonfam,\phonl,\phonm,\phonn)$, where $\phonl \in \N_0$, $\phonm \in \Z, |\phonm|\leq \phonl$, and $\phonn \in \N$. The radial order $\phonn$, polar order $\phonl$, and azimuthal order $\phonm$ count the number of nodes of the mode function $\wmode_\phonindex(\pos)$ in the direction of the three spherical coordinates $(\rpos,\polpos,\azpos)$, respectively, and we use the terms $\Tmode_{\phonl\phonm\phonn}$ or $\Smode_{\phonl\phonm\phonn}$ to name each eigenmode. The corresponding displacement modal fields $\wmode_\phonindex(\pos)$ can be calculated analytically and are given in \cref{tab: phonon displacement modal fields} in \cref{sec: sphere appendix}. Based on these results we obtain the explicit expressions for $\linshift_\phonindex$, $\bscoupling_{\phonindex\phonindexb}$, and $\sqcoupling_{\phonindex\phonindexb}$ listed in \cref{tab: coupling constants sphere} which we later use in the case study in \cref{sec: case study}.

\subsection{Shape and Acoustic Properties}
\label{sec: acoustic properties}

The shape and acoustic properties of a spinning nanoparticle can be inferred from the Hamiltonian \cref{eqn: fixed rotation total Hamiltonian} and are affected in two ways: First, the static centrifugal force $\dens \rotfreq^2 \pos_\perp$ changes the equilibrium mass distribution and hence the shape of the nanoparticle. Second, the dynamical centrifugal force $\dens \rotfreq^2 \ufield_\perp(\post)$ modifies both the spatial profile and the frequencies of the phononic eigenmodes.

The new shape is described by a static contribution $\ufield_0(\pos)$ to the displacement field $\ufield(\post) = \ufield_0(\pos) + \displaced{\ufield}(\post)$. The remaining dynamical part $\displaced{\ufield}(\post)$ represents  vibrations around this new equilibrium configuration. On the level of the quantum theory, $\ufield_0(\pos)$ can be determined via a unitary transformation that cancels the linear Hamiltonian $\HamilopIntLinear$ and displaces the ladder operators $\phonaop_\phonindex = \phondispl_\phonindex + \displaced{\phonaop}_\phonindex$ by complex numbers $\phondispl_\phonindex$ \footnote{The shift of the ladder operators $\phonaop_\phonindex$ by complex numbers  $\phondispl_\phonindex$ corresponds to a unitary transformation with the mode displacement operator $\displop \equiv \bigotimes_\phonindex \exp ( \phondispl_\phonindex \hconj\phonaop_\phonindex - \cconj\phondispl_\phonindex \phonaop_\phonindex)$ such that any operator $\op{O}$ is transformed as $\op{O}^d \equiv \hconj\displop \op{O} \displop$. Subsequently, we mark operators in the initial representation with an underline and drop the index $d$ denoting the new representation.}. The displacement field operator is then
\begin{equation}
  \ufieldop(\pos) = \ufield_0(\pos) + \displaced{\ufieldop}(\pos)
\end{equation}
where the dynamical part $\displaced{\ufieldop}(\pos)$ is of the form \cref{eqn: displacement field operator} but with displaced ladder operators $\displaced{\phonaop}_\phonindex$ replacing the ladder operators $\phonaop_\phonindex$. The static part is
\begin{equation}\label{eqn: static displacement field}
     \ufield_0(\pos) = 2 \sum_\phonindex \ufieldmodedens_\phonindex \real \spare{ \phondispl_\phonindex \,\wmode_\phonindex(\pos)},
\end{equation}
where the mode displacements $\phondispl_\phonindex$ need to satisfy
\begin{equation}\label{eqn: condition for mode displacements}
  \sum_\phonindexb [ \kronecker_{\phonindex\phonindexb}\phonfreq_\phonindexb \cconj\phondispl_\phonindexb + \cconj{\bscoupling}_{\phonindex\phonindexb} \cconj{\phondispl}_\phonindexb +  \sqcoupling_{\phonindex\phonindexb} \phondispl_\phonindexb  ]= - \linshift_\phonindex
\end{equation}
such that the displaced Hamiltonian
\begin{equation}\label{eqn: quadratic Hamiltonian}
  \Hamilop = \HamilopBareDisplaced + \HamilopIntQuadraticDisplaced
\end{equation}
is purely quadratic. Here, we drop a constant energy term. The bare Hamiltonian $\HamilopBareDisplaced$ and hybridization Hamiltonian $\HamilopIntQuadraticDisplaced$ are defined as in \cref{eqn: fixed rotation total Hamiltonian contributions 2} but substituting the ladder operators $\phonaop_\phonindex$ with the displaced operators $\displaced{\phonaop}_\phonindex$.
\latin{A priori}, there is an infinite number of coupled conditions \cref{eqn: condition for mode displacements}, precluding the direct computation of the $\phondispl_\phonindex$. However, it is possible to approximate $\ufield_0(\pos)$ by discarding all but a sufficient number $N$ of low frequency modes because the constants $\linshift_\phonindex$, $\bscoupling_{\phonindex\phonindexb}$, and $\sqcoupling_{\phonindex\phonindexb}$ tend to zero at high phonon frequencies. In performing this truncation, $N$ needs to be chosen sufficiently large to ensure that the most relevant displacements of the lowest-frequency modes are well approximated; see \cref{sec: case study} for details. The truncation reduces \cref{eqn: condition for mode displacements} to a system of $2N$ real-valued linear equations for the $N$ real parts and $N$ imaginary parts of the mode displacements $\phondispl_\phonindex$ that can be solved directly. Note that the expectation value of the displacement field in a thermal state $\braket{\displaced{\ufieldop}}_\thermal(\pos) = \ufield_0(\pos)$ reflects the fact that the static field describes the new equilibrium shape of the rotating nanoparticle. By construction, the static field $\ufield_0(\pos)$ balances the elastic restoring force and the static centrifugal force, $\Dphon \ufield_0(\pos) + \dens \rotfreq^2 \pos_\perp + \dens \rotfreq^2 {\ufield_0}_\perp(\pos) = 0$ such that the equation of motion simplifies to $\dens \ddeltt{\displaced{\ufield}}(\post) = \Dphon \displaced{\ufield}(\post) + \dens \rotfreq^2 \displaced{\ufield}_\perp(\post)$.

Let us now focus on the acoustic properties of a spinning nanoparticle. The phonon eigenfrequencies are reduced because the dynamical centrifugal force acts in a direction opposed to the elastic restoring force. Moreover, the spatial shape of the vibrational eigenmodes is modified. Since the coupling constants $\bscoupling_{\phonindex\phonindexb}$ and $\sqcoupling_{\phonindex\phonindexb}$ decay with increasing frequency $\phonfreq_\phonindex$, we can again focus on the set of $N$ modes $\phonindex$ of the resting nanoparticle that are lowest in frequency in order to approximate the $N$ lowest eigenmodes of the spinning nanoparticle. We construct these new phononic eigenmodes by diagonalizing the quadratic Hamiltonian \cref{eqn: quadratic Hamiltonian} via a Bogoliubov transformation~\cite{bogoljubov_new_1958,valatin_comments_1958}; see \cref{sec: sphere appendix} for details. It is possible to diagonalize a finite-dimensional quadratic Hamiltonian in terms of bosonic modes if and only if it is linearly stable, that is, if it does not lead to the divergence of any observable over time; see, e.g., Refs.~\cite{rusconi_quantum_2017,kustura_quadratic_2019}. In the present situation this is the case provided the dynamical centrifugal force is not so large as to reduce any phonon eigenfrequency to zero. In consequence, there is a critical rotation frequency $\rotfreqcrit$ above which the Hamiltonian \cref{eqn: quadratic Hamiltonian} is linearly unstable. Note however that anharmonic contributions to the interatomic interaction already need to be accounted for when approaching this critical frequency. In \cref{sec: nonlinearity appendix}, we derive a criterion that allows us to define a rotation frequency $\rotfreqnl<\rotfreqcrit$ where nonlinear elastic effects start to become relevant. Hence, for rotational frequencies $\rotfreq > \rotfreqnl$ one should describe the system beyond linear elastodynamics which is an interesting further research direction that we do not investigate in this manuscript. Within linear elastodynamics and in the linearly stable regime $\rotfreq < \rotfreqcrit$, we can express the Hamiltonian in its diagonal normal form
\begin{equation}
 \Hamilop = \hbar \sum_{\phonindexInt} \phonfreqInt_\phonindexInt \hconj\phonaopInt_\phonindexInt \phonaopInt_\phonindexInt
\end{equation}
where the index $\phonindexInt$ labels the new eigenmodes, $\phonaopInt_\phonindexInt$ are the corresponding ladder operators, and $\phonfreqInt_\phonindexInt$ are the new eigenfrequencies. It is then possible to construct the displacement modal fields $\wmodeInt_\phonindexInt(\pos)$ of the new eigenmodes such that the dynamical part of the displacement field can be expanded as
\begin{equation}
  \displaced{\ufieldop} (\pos) = \sum_\phonindexInt \ufieldmodedens_\phonindexInt [\phonaopInt_\phonindexInt \wmodeInt_\phonindexInt(\pos)+ \hc ]
\end{equation}
with the mode density $\ufieldmodedens_\phonindexInt \equiv \sqrt{ \hbar/ (2\dens \phonfreqInt_\phonindexInt)}$; see \cref{sec: sphere appendix} for details. By construction, the fields $\wmodeInt_\phonindexInt(\pos)$ are eigenfunctions of the differential operator $\DphonInt$ that includes both the restoring force and the dynamical centrifugal force $\DphonInt\displaced{\ufield}(\post) = \Dphon \displaced{\ufield}(\post) + \dens \rotfreq^2 \displaced{\ufield}_\perp(\post)$.

\subsection{Optical Properties of a Subwavelength Sphere}
\label{sec: optical properties}

We now turn to analyzing the optical properties of a spinning nanoparticle. The changes in its equilibrium configuration affect the electric permittivity and polarizability (even in the absence of phonons). In particular, we consider a nonabsorbing and nonmagnetic dielectric nanoparticle that is optically homogeneous and isotropic with a real valued relative permittivity $\relpermitt$ while at rest. We assume $\relpermitt$ to be constant over the relevant range of wavelengths of light.
A spinning nanoparticle is no longer homogeneous nor isotropic due to the photoelastic effect~\cite{narasimhamurty_photoelastic_1981,nelson_theory_1971}: The permittivity is modulated locally by variations in the mass distribution and the centrifugal force induces a strain $\straintens_0(\pos)$ that introduces local optical axes. In consequence, the optical properties need to be described by a position-dependent permittivity tensor $\relpermitttensb(\pos)$. Since a nanoparticle is smaller than the wavelengths of light, it is useful to consider the effective permittivity $\relpermitttensb$ obtained from the strain $\averageStrain \equiv \int_\body \straintens_0(\pos) \dd\pos / V$ averaged over the volume $V$ of the nanoparticle. The permittivity tensor is then constant and diagonal with diagonal elements
\begin{equation}\label{eqn: relative permittivity of spinning sphere}
    {\relpermitttensbcomp}^{ii} = \frac{\relpermitt}{1+\relpermitt \Delta\impermitttenscomp^{ii} },
\end{equation}
where $i\in{\xpos,\ypos,\zpos}$. The strain-induced correction $\Delta \impermitttenscomp^{ij} \equiv \pockelstenscomp^{ijkl} \averageStrainComp^{kl}$ to the inverse permittivity is quantified heuristically using the dimensionless photoelasticity tensor $\pockelstens$~\cite{narasimhamurty_photoelastic_1981,nelson_theory_1971}. For isotropic homogeneous solids, the photoelasticity tensor has only two independent components $\photela$, $\photelb$ and
\begin{equation}\label{eqn: change in impermittivity}
  \Delta \impermitttenscomp^{ii} = \photela \averageStrainComp^{ii} + \photelb \sum_{j\neq i}  \averageStrainComp^{jj};
\end{equation}
see Ref.~\cite{narasimhamurty_photoelastic_1981} for details. The average strain due to rotation is
\begin{equation}\label{eqn: mode expansion average strain}
    \averageStrain = 2\sum_\phonindex \ufieldmodedens_\phonindex \real\spare{ \phondispl_\phonindex \averageStrainMode_\phonindex(\pos) },
\end{equation}
where $\averageStrainMode_\phonindex(\pos)$ are the averages of the strain modal fields related to the displacement modal fields $\wmode_\phonindex(\pos)$ analogous to \cref{eqn: definition strain and stress tensor}.

Let us further assume that the nanoparticle is spherical with radius $\rad$ such that it is deformed into an oblate spheroid under rotation around the $\zpos$ axis; compare \cref{sec: case study}. The polarizability of a homogeneous and isotropic sphere is given by $\polarizab = 3\vacpermitt V (\relpermitt-1)/(\relpermitt+2)$ within the dipole approximation~\cite{bohren_absorption_2008,hulst_light_2012}. If the nanoparticle is spinning, its permittivity, aspect ratio, and volume change. In consequence, its optical response is no longer isotropic. However, the response field can still be calculated analytically for the case of a homogeneous ellipsoid and a tensor-valued polarizability $\polarizabtens$ can be defined. The polarizability tensor is diagonal with elements~\cite{bohren_absorption_2008,hulst_light_2012}
\begin{equation}\label{eqn: polarizability of spinning sphere}
    \polarizab^{ii} = 3\vacpermitt \volumeEllipsoid \frac{\relpermitttensbcomp^{ii}-1}{3 + 3 \geometricFactor_{i} ( \relpermitttensbcomp^{ii}-1)}.
\end{equation}
Here, $\volumeEllipsoid = 4\pi \halfAxisA \halfAxisB \halfAxisC/3$ is the volume of the ellipsoid and $\halfAxis_i$ are the lengths of its half axes. For an oblate spheroid, the geometric factors $\geometricFactor_i$ are~\cite{bohren_absorption_2008,hulst_light_2012}
\begin{equation}
  \begin{split}
  \geometricFactorA &= \geometricFactorB = \frac{g(e)}{2e^2} \spare{ \frac{\pi}{2} - \atan g(e) } - \frac{g^2(e)}{2}, \\
  \geometricFactorC &= 1 - 2 \geometricFactorA,
  \end{split}
\end{equation}
where $g(e) \equiv \sqrt{(1-e^2)/e^2}$ is a function of the eccentricity $e \equiv \sqrt{ 1 - \halfAxisC^2/\halfAxisA^2}$. The lengths of the half axes $\halfAxis_i$ are obtained by adding the static displacement \cref{eqn: static displacement field} on the surface to the nanosphere radius. By using the permittivity \cref{eqn: relative permittivity of spinning sphere} in \cref{eqn: polarizability of spinning sphere}, one thus obtains the polarizability for a spinning nanosphere.

\section{Results for a Dielectric Nanosphere}
\label{sec: case study}

\begin{table}
  \newcolumntype{A}{>{\begin{math}}r<{\end{math}}}
  \newcolumntype{B}{>{\begin{math}}c<{\end{math}}}
  \newcolumntype{C}{>{\begin{math}}l<{\end{math}}}
  \newcolumntype{R}{>{\raggedleft\arraybackslash}X<{}}
  \begin{tabularx}{\columnwidth}{ABC @{\quad} l R}
    \toprule
    \multicolumn{3}{l}{Independent parameters}\\
      \rad &=& \SI{100}{\nano\meter} & Sphere radius &~\cite{reimann_ghz_2018} \\
      \dens &=& \SI{2.20}{\gram/\centi\meter^3} & Mass density &~\cite{bass_handbook_2001} \\
      \lamelambda &=& \SI{15.2}{\giga\pascal} & Linear elastic constants &~\cite{bass_handbook_2001}\\
      \lamemu &=& \SI{31.2}{\giga\pascal} & \\
      \anharmonicA &=& \SI{-44}{\giga\pascal} & Nonlinear elastic constants &~\cite{bogardus_thirdorder_1965}\\
      \anharmonicB &=& \SI{93}{\giga\pascal} & \\
      \anharmonicC &=& \SI{28}{\giga\pascal} & \\
      \relpermitt &=& \num{2.1} & Relative permittivity &~\cite{bass_handbook_2001} \\
      \photela &=& \num{0.100} & Photoelasticity coefficients &~\cite{vedam_elastic_1950} \\
      \photelb &=& \num{0.285} &  \\
      \midrule
      \multicolumn{3}{l}{Derived parameters} \\
      \clong &=& \SI{5.94e3}{\meter/\second} & Longitudinal sound speed \\
      \ctrans &=& \SI{3.76e3}{\meter/\second} & Transverse sound speed \\
      \Poissonnu &=& \num{0.164} & Poisson ratio\\
    \bottomrule
  \end{tabularx}
  \caption{Physical parameters used in the case study based on the experiment on optically levitated silica nanospheres reported in Ref.~\cite{reimann_ghz_2018}. The elastic constants $\lamelambda$ and $\lamemu$ together with the mass density $\dens$ of the nanoparticle describe its linear acoustic properties. In particular, they determine the Poisson ratio $\Poissonnu \equiv \lamelambda/[2 (\lamelambda +\lamemu )]$ that relates the elasticities under compression and shear~\cite{gurtin_linear_1984}, as well as the longitudinal and transverse sound speeds $\clong$ and $\ctrans$; see \cref{eqn: definition sound speeds}. The optical properties are determined by the relative permittivity $\relpermitt$ and the components $\photela$, $\photelb$ of the photoelasticity tensor.}
  \label{tab: parameters}
\end{table}

\begin{figure*}
  \includegraphics[height=185.46pt]{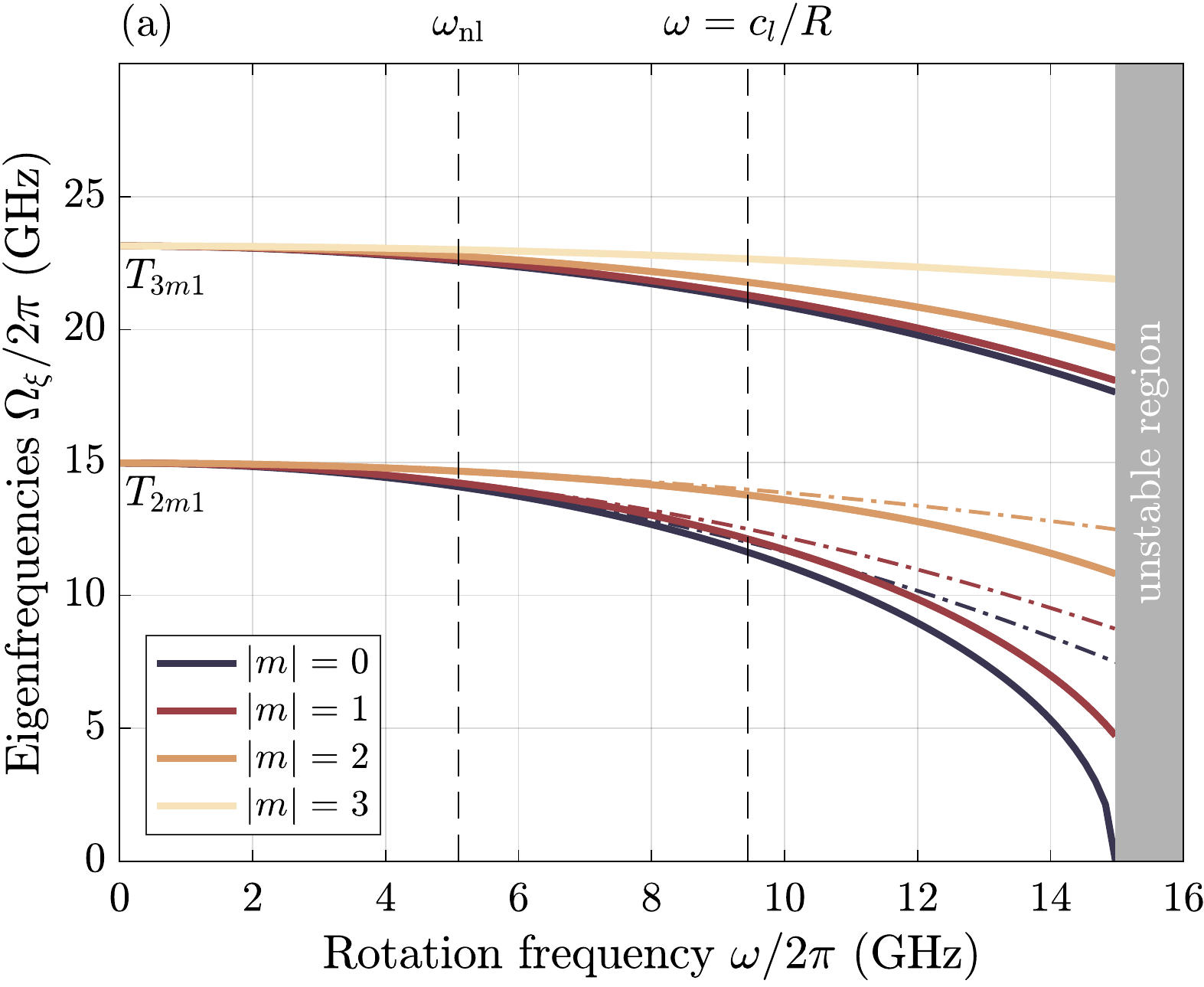}
  \includegraphics[height=185.46pt]{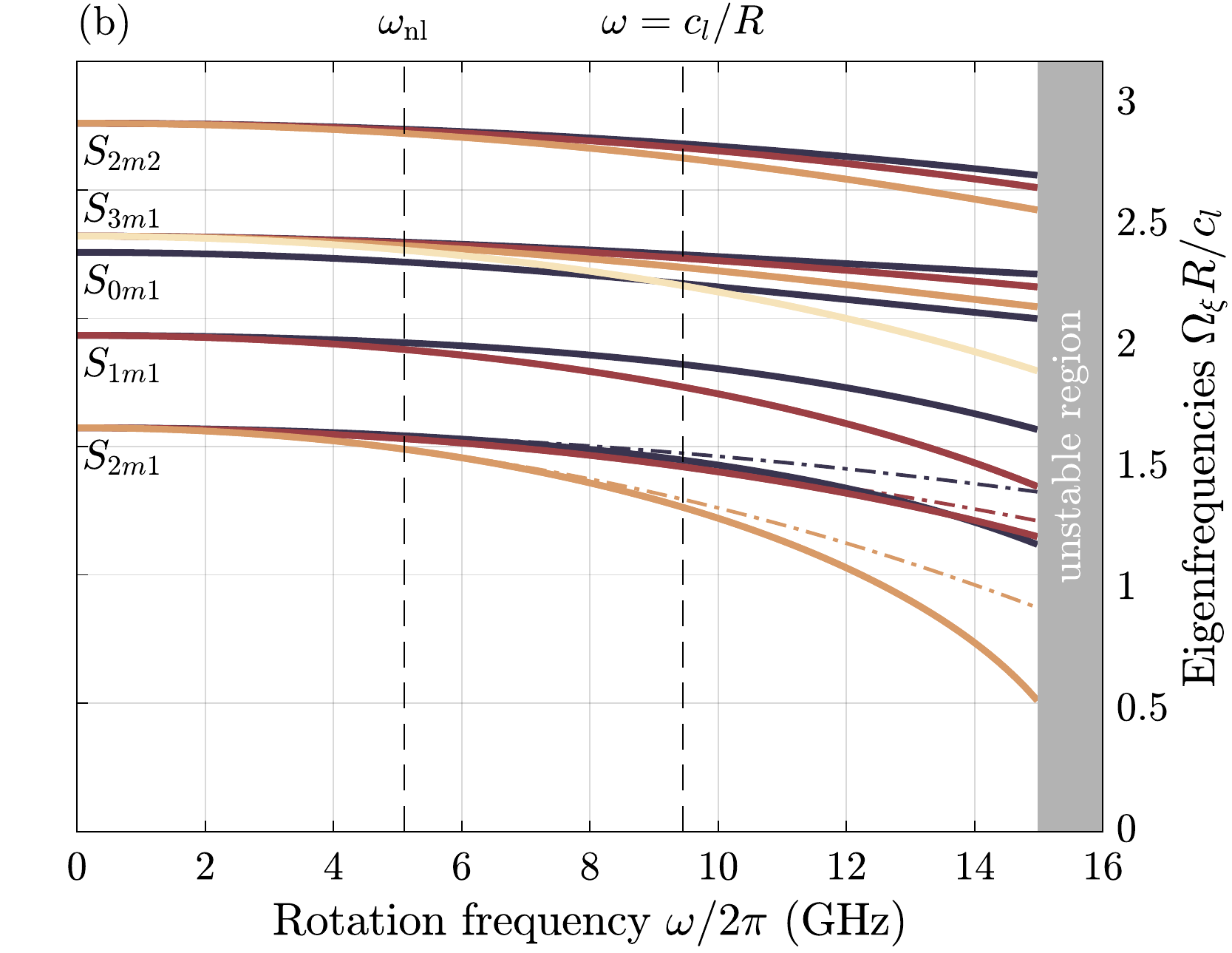}
  \caption{Phononic eigenfrequency spectrum of a spinning silica nanosphere as a function of its rotation frequency. Panel~(a) shows the spectrum of torsional modes and panel~(b) the spectrum of spheroidal modes. The parameters of the nanosphere are specified in \cref{tab: parameters}. The labels $\Tmode_{\phonl\phonm\phonn}$ and $\Smode_{\phonl\phonm\phonn}$ and the line colors indicate from which bare mode $\phonindex$ of the resting sphere each hybridized mode $\phonindexInt$ of the spinning sphere originates. While there is a degeneracy in the azimuthal order $\phonm$ at rest, the degeneracy is partially lifted by the rotation. At rotation frequencies $\rotfreq \ll \clong/\rad$, the shift in eigenfrequencies scales as $\phonfreq_\phonindex - \phonfreqInt_\phonindexInt \propto \rotfreq^2 \rad/ \clong$ as indicated by the dashed-dotted lines. Elastic nonlinearities will first affect the $\Smode_{001}$ and $\Smode_{201}$ modes at frequencies $\rotfreq \geq \rotfreqnl$ and can modify the entire spectrum at frequencies starting around $\rotfreq \simeq \clong/\rad$. Note that there are crossings of eigenfrequencies without leading to strong hybridization between the modes. The linearly unstable region $\rotfreq\geq\rotfreqcrit$ begins where the calculated frequency of the fundamental mode originating from the $\Tmode_{201}$ mode drops to zero.}
  \label{fig: phonon spectrum spinning sphere}
\end{figure*}

We now explicitly consider a homogeneous and isotropic silica nanosphere of radius $\rad = \SI{100}{\nano\meter}$ with parameters specified in \cref{tab: parameters}. The rotation of such nanoparticles at $\si{\giga\hertz}$ frequencies has recently been reported in Ref.~\cite{reimann_ghz_2018}. The linear acoustic properties of the nanoparticle are described by the elastic constants $\lamelambda$ and $\lamemu$ together with the mass density $\dens$; see \cref{sec: elastodynamics appendix}. The optical properties are determined by the relative permittivity $\relpermitt$ and the components $\photela$, $\photelb$ of the photoelasticity tensor. In the following analysis, we consider all phonon modes with polar order $\phonl \leq 3$ and radial order $\phonn \leq 3$. We choose this truncation limit because it is sufficient to ensure convergence to the third relevant digit in quantities like the size of the nanoparticle. Including higher order modes is straightforward but only leads to negligible corrections for the results presented here~%
\footnote{It is convenient to include only phonon modes with polar order $\phonl\leq\phonl_\max$ and radial order $\phonn \leq \phonn_\max$ which amounts to $N=2\phonn_\max(\phonl_\max+1)^2-3$ modes. The results presented in \cref{sec: case study} were obtained for $(\phonl_\max, \phonn_\max, N) = (3,3,93)$. Including higher order phonon modes has no relevant impact on our findings. First of all, the displacements drop quickly with $\phonn$. For instance, the displacements of the $\Smode_{003}$ and $\Smode_{203}$ modes are more than two orders of magnitude smaller than the displacements of the $\Smode_{001}$ and $\Smode_{201}$ modes. The static field $\ufield_0(\pos)$ as well as the optical properties we infer from it are therefore numerically stable \latin{vis-à-vis} the trunctation of \cref{eqn: condition for mode displacements}: Including higher order modes up to $(\phonl_\max,\phonn_\max,N)=(5,5,357)$, for example, modifies the prediction for the change of the half axis $\halfAxisA-\rad$ by only $\SI{0.4}{\percent}$ at $\rotfreq = 2\pi \times \SI{5}{\giga\hertz}$. Similarly, the part of the phonon spectrum of the spinning particle shown in \cref{fig: phonon spectrum spinning sphere} converges quickly when increasing the number of modes included. Indicative is that the constant $\relHybridizationCritical$ characterizing the linear stability the Hamiltonian \cref{eqn: quadratic Hamiltonian} only changes by about $\SI{0.5}{\percent}$ when extending the analysis to $357$ phonon modes.}.

\begin{figure}
  \begin{centering}
    \includegraphics[width=234.66pt]{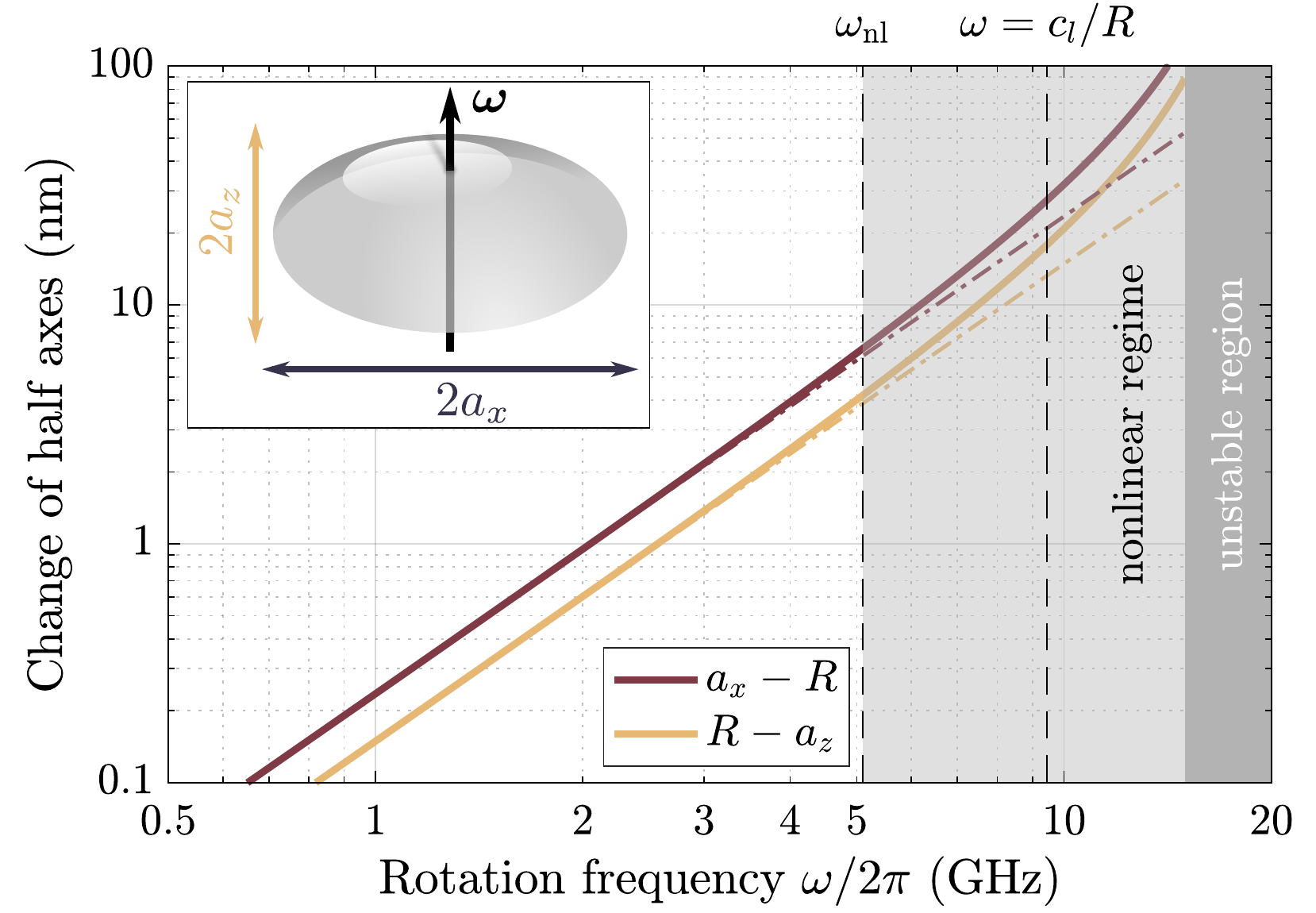}
  \end{centering}
  \caption{Change of shape of a spinning silica nanosphere of radius $\rad=\SI{100}{\nano\meter}$ with elastic properties specified in \cref{tab: parameters}. The solid lines are obtained from \cref{eqn: static displacement field,eqn: condition for mode displacements} and indicate the increase of the half axes $\halfAxisA=\halfAxisB$ in the equatorial plane and the decrease of the half axis $\halfAxisC$ in axial direction. In the regime $\rotfreq \ll \clong/\rad$, the change in shape is proportional to $\rotfreq^2\rad^3/\clong^2$ as indicated by the dashed-dotted lines. The light gray area indicates the regime $\rotfreq \geq \rotfreqnl$ where anharmonic corrections to the linear elastic theory used here become increasingly relevant; see \cref{sec: nonlinearity appendix}. The dark gray area indicates the linearly unstable region $\rotfreq \geq \rotfreqcrit$.}
  \label{fig: change of shape}
\end{figure}

A spinning nanosphere is deformed into an oblate spheroid \footnote{The spheroidal shape of a spinning nanosphere can for instance be verified numerically by checking that the size $R+\ufield_0(\pos)$ of the nanoparticle obeys the equation of an ellipsoid.}. \Cref{fig: change of shape} shows the predicted change in the shape of the nanosphere. The solid lines indicate the length of the equatorial half axes $\halfAxisA = \halfAxisB$ and the axial half axis $\halfAxisC$ as a function of the rotation frequency $\rotfreq$. They are obtained by evaluating the static displacement field $\ufield_0(\pos)$ defined in \cref{eqn: static displacement field}. For reasons of symmetry, only the $\Smode_{00\phonn}$ and $\Smode_{20\phonn}$ modes have nonzero displacements $\phondispl_\phonindex$ and contribute to $\ufield_0(\pos)$. In the weak hybridization limit $\rotfreq \ll \clong/\rad$, the change in shape follows a power law. The eigenfrequencies of the phonon modes of a nanosphere scale as $\phonfreq_\phonindex \propto \clong/\rad$ and the coupling frequencies as $\bscoupling_{\phonindex\phonindexb},\sqcoupling_{\phonindex\phonindexb} \propto \rotfreq^2\rad/\clong$ where $\clong = \sqrt{(2\lamemu+\lamelambda)/\dens}$ is the longitudinal sound speed; see \cref{sec: sphere appendix}. Therefore $\bscoupling_{\phonindex\phonindexb},\sqcoupling_{\phonindex\phonindexb} \ll \phonfreq_\phonindex$ at low rotation frequencies and the Eqs.~(\ref{eqn: condition for mode displacements}) decouple. The mode displacements are then $\phondispl_\phonindex \simeq - \cconj\linshift_\phonindex/\phonfreq_\phonindex$ and the change in shape scales as
\begin{equation}
\ufield_0(\pos) \propto \rotfreq^2 \rad^3/\clong^2.
\end{equation}
This power-law dependence is indicated by the dashed-dotted lines in \cref{fig: change of shape}. Beyond rotation frequencies of $\rotfreq \simeq \clong/\rad$, the mode hybridization starts to become relevant and the linear elastic theory predicts deviations from this power law. For the particular case of fused silica, we expect anharmonicities in the interatomic interaction potential to appear at similar frequencies: In \cref{sec: nonlinearity appendix}, we define the rotational frequency $\rotfreqnl$ at which anharmonic contributions increase the elastic energy of the displaced $\Smode_{001}$ mode by $\SI{25}{\percent}$. For fused silica, $\rotfreqnl = 2\pi\times\SI{5.1}{\giga\hertz}$. In the regime $\rotfreq \geq \rotfreqnl$, indicated by the light gray area in \cref{fig: change of shape}, nonlinear corrections to the linear elastic theory presented in this work may thus become sizable and increasing deviations from the shape predicted by the linear theory are to be expected in practice. Finally, the dark gray area marks the region at rotation frequencies greater than $\rotfreqcrit$ for which the Hamiltonian \cref{eqn: fixed rotation total Hamiltonian} is linearly unstable and linearized elasticity theory is no longer applicable as we discuss in \cref{sec: theory}. For a spherical nanoparticle, the critical frequency scales as $\rotfreqcrit^2 \rad^2/\clong^2 = \relHybridizationCritical$ where $\relHybridizationCritical$ is a constant that depends only on the Poisson ratio $\Poissonnu \equiv \lamelambda/[2 (\lamelambda +\lamemu )]$. In the case of fused silica, $\relHybridizationCritical = 2.5$ such that $\rotfreqcrit = 2\pi \times \SI{15}{\giga\hertz}$ in this case study. At a rotation frequency of $\rotfreq=2\pi \times \SI{5}{\giga\hertz}$, our linear elastic theory predicts the equatorial diameters to change by $2(\halfAxisA - \rad) \simeq \SI{12}{\nano\meter}$ and  $2(\rad -\halfAxisC)\simeq \SI{8}{\nano\meter}$. Such frequencies can for instance be achieved with levitated nanodumbbells consisting of two silica nanospheres~\cite{ahn_ultrasensitive_2020}.

\begin{figure*}[ht!]
  \begin{centering}
    \includegraphics[width=228.8823pt]{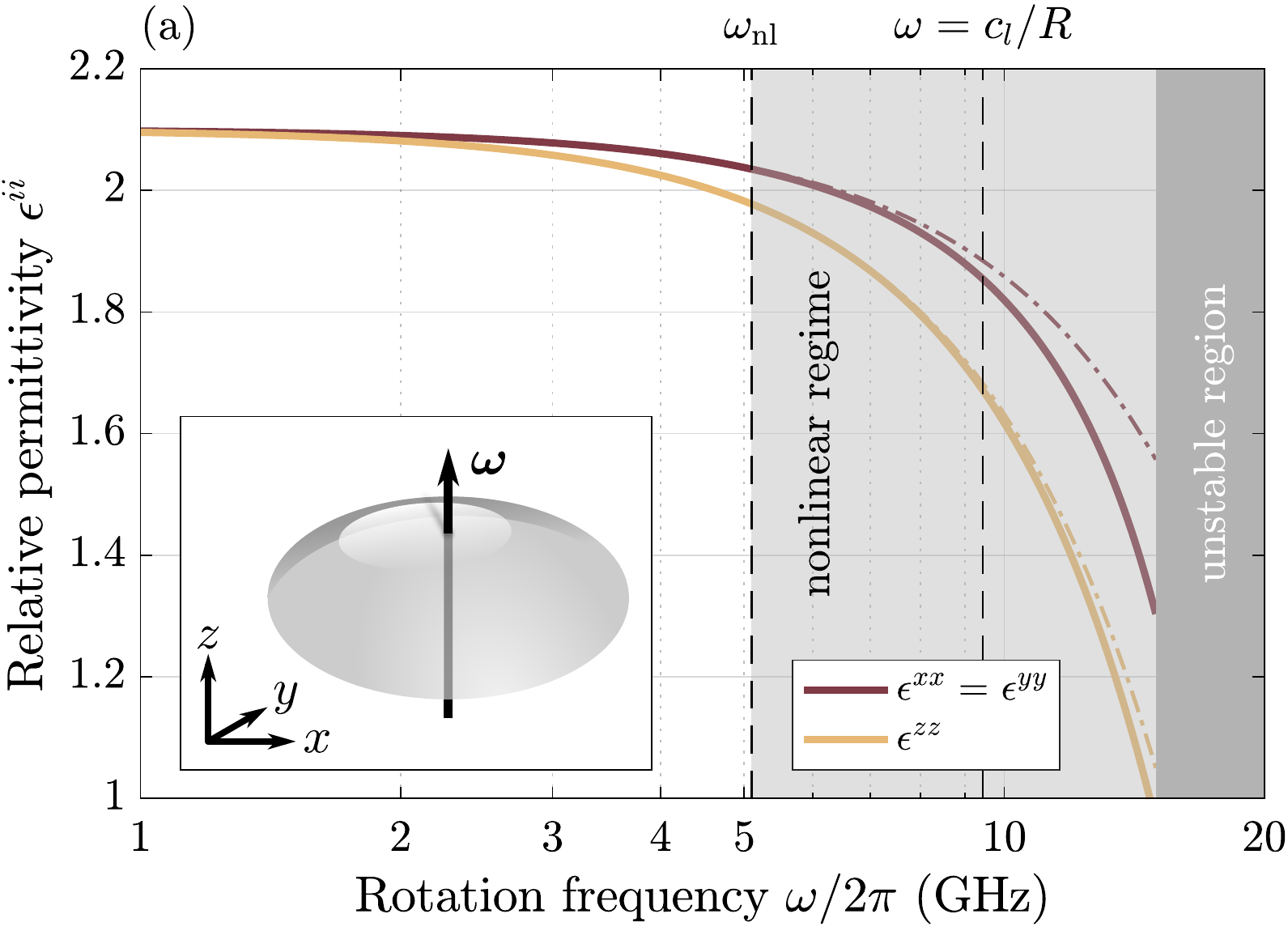}
    \includegraphics[width=261.6601pt]{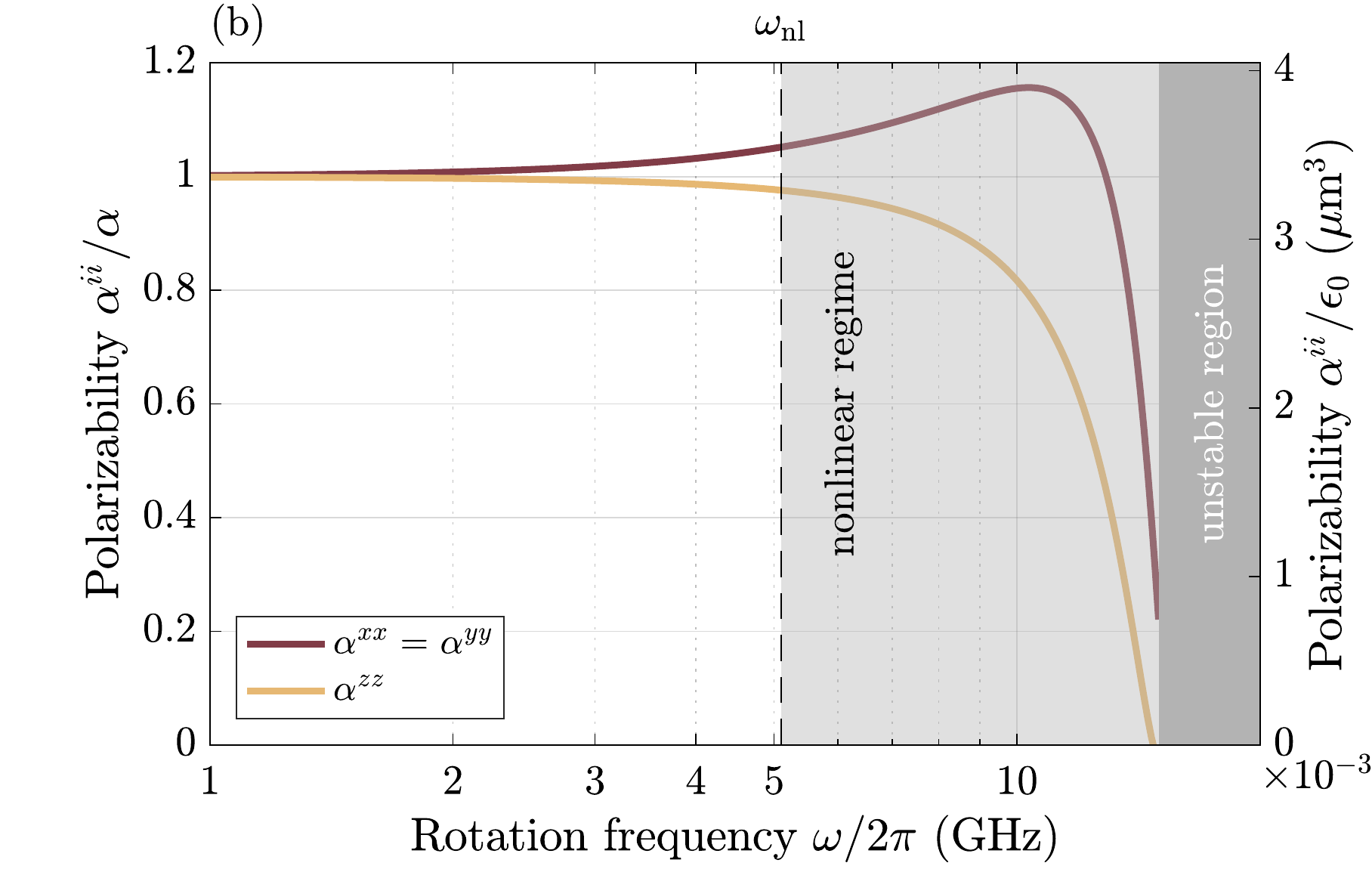}
    \caption{Optical properties of a spinning silica nanosphere depending on its rotation frequency. The parameters of the nanosphere are specified in \cref{tab: parameters}. The solid lines in panel~(a) represent the components of the permittivity tensor obtained from \cref{eqn: relative permittivity of spinning sphere} and demonstrate that the rotation causes birefringence $ \relpermitttensbcomp^{\zpos\zpos} \neq \relpermitttensbcomp^\xpos\xpos =  \relpermitttensbcomp^{\ypos\ypos}$. At low rotation frequencies $\rotfreq\ll\clong/\rad$, the change in the permittivity is proportional to $\relpermitt^2\rotfreq^2\rad^2/\clong^2$ as indicated by the dashed-dotted lines. Panel~(b) shows the components of the polarizability tensor calculated using \cref{eqn: polarizability of spinning sphere}. Here, $\polarizab$ is the polarizability of the resting sphere. Note that in the regime $\rotfreq \geq \rotfreqnl$ (see \cref{sec: nonlinearity appendix})  indicated by the light gray area, nonlinear effects are increasingly relevant for the actual shape of the nanoparticle and deviations from the predictions of our linear elastic theory are to be expected in practice.}
    \label{fig: permittivity and polarizability}
  \end{centering}
\end{figure*}

In \cref{fig: phonon spectrum spinning sphere}, we plot the dependence of the phonon eigenfrequency spectrum $\phonfreqInt_\phonindexInt$ on the rotation frequency. The labels $\Tmode_{\phonl\phonm\phonn}$ and $\Smode_{\phonl\phonm\phonn}$ indicate from which bare eigenmode $\phonindex$ of the resting sphere each hybridized eigenmode $\phonindexInt$ of the spinning sphere originates. For reference, the eigenfrequencies $\phonfreq_\phonindex$ of a resting sphere are shown in \cref{fig: sphere phonon spectrum} in \cref{sec: sphere appendix}. At rest, the spectrum is degenerate in the azimuthal order $\phonm$. This degeneracy is partially lifted at $\rotfreq>0$, leaving only modes $\pm\phonm$ degenerate. The color of each spectrum line indicates the magnitude $|\phonm|$ of the polar order of the bare mode from which it originates. In the limit $\rotfreq \ll \clong/\rad$ of a slowly spinning nanoparticle,
the spatial shape of every hybridized mode $\phonindexInt$ closely resembles the shape of a bare mode $\phonindex$ of the resting nanoparticle $\wmodeInt_\phonindexInt(\pos) \simeq \wmode_\phonindex(\pos)$. We can then approximate the change of the eigenfrequency of each mode $\phonindexInt\sim\phonindex$ by considering  the direct frequency shifts in the Hamiltonian \cref{eqn: quadratic Hamiltonian} only (in first-order perturbation): $\phonfreqInt_\phonindexInt \simeq \phonfreq_\phonindex + \bscoupling_{\phonindex\phonindex}$. The constants $\bscoupling_{\phonindex\phonindex}$ are negative, consistent with a reduction of the eigenfrequencies $\phonfreqInt_\phonindex < \phonfreq_\phonindex$ and the shift in eigenfrequencies scales as
\begin{equation}
 \phonfreq_\phonindex - \phonfreqInt_\phonindexInt \propto \frac{\rotfreq^2 \rad}{\clong}.
\end{equation}
Moreover, since the constants $\bscoupling_{\phonindex\phonindex}$ tend to zero at high frequencies $\phonfreq_\phonindex$, modes of lower frequencies are more strongly affected. Nonlinear effects become relevant for the $\Smode_{00\phonn}$ and $\Smode_{20\phonn}$ modes displaced by the static centrifugal force at $\rotfreq\geq \rotfreqnl$; see \cref{sec: nonlinearity appendix}. Other modes are affected once they hybridize with the displaced modes at frequencies beyond $\rotfreq \simeq \clong/\rad$. The onset of linear instability is marked by the reduction of the eigenfrequency of the lowest-frequency phonon mode originating from the $\Tmode_{201}$ mode to zero.

At a rotation frequency of $\rotfreq=2\pi \times \SI{5}{\giga\hertz}$, the frequency of the $\Smode_{221}$ mode is reduced by $(\phonfreq_\phonindex - \phonfreqInt_\phonindexInt)/2\pi \simeq \SI{0.4}{\giga\hertz}$. In experimental studies of Brillouin scattering off silica nanospheres on a substrate, this mode was found to be optically active with a linewidth on the order of few $\si{\giga\hertz}$~\cite{kuok_brillouin_2003}. Since the phonon linewidths in levitated nanospheres are expected to be lower, the rotation-induced shift in the phonon spectrum and possibly even nonlinear corrections to the predictions presented in \cref{fig: phonon spectrum spinning sphere} should be measurable.

Let us now discuss the optical properties. \Cref{fig: permittivity and polarizability} shows the permittivity $\relpermitttensb$ and polarizability $\polarizabtens$ of the spinning nanosphere as functions of the rotation frequency. Both tensors are diagonal as discussed in \cref{sec: optical properties}. In panel~(a) of \cref{fig: permittivity and polarizability}, we plot the diagonal elements $\relpermitttensbcomp^{ii}$ of the permittivity tensor. The solid lines correspond to the values predicted by \cref{eqn: relative permittivity of spinning sphere} with the average strain of the relevant $\Smode_{00\phonn}$ and $\Smode_{20\phonn}$ modes given in \cref{eqn: average strain S00n modes,eqn: average strain S2mn modes}. The permittivity is decreasing with increasing rotation frequency, and there is an increasing birefringence $\relpermitttensbcomp^{\zpos\zpos} \neq \relpermitttensbcomp^{\xpos\xpos} = \relpermitttensbcomp^{\ypos\ypos}$ between the axial direction and the equatorial plane. In the limit $\rotfreq \ll \clong/\rad$ of a slowly spinning nanoparticle, the change in the permittivity scales as
\begin{equation}
  \relpermitt - {\relpermitttensbcomp}^{ii}  \propto \relpermitt^2 \frac{\rotfreq^2\rad^2}{\clong^2},
\end{equation}
indicated by the dashed-dotted lines in \cref{fig: permittivity and polarizability}. In panel~(b) of \cref{fig: permittivity and polarizability}, we plot the diagonal elements $\polarizab^{ii}$ of the polarizability tensor calculated according to \cref{eqn: polarizability of spinning sphere}. The frequency-dependence of the polarizability is caused by the changes in the permittivity, the aspect ratio, and the volume of the nanoparticle. The increase of the polarizability in equatorial direction and its decrease in axial direction are driven by the increasing aspect ratio. The subsequent drop in $\relpermitttensbcomp^{\xpos\xpos}$ and $\relpermitttensbcomp^{\ypos\ypos}$ is due to the decreasing volume of a fast spinning ellipsoidal nanoparticle with a large aspect ratio. Note that for the case of fused silica, this behavior is observed in the regime $\rotfreq \geq \rotfreqnl$ where nonlinear elastic effects need to be accounted for. However, more elastic materials like polymers or even liquid helium could still exhibit such a behavior in the linear elastic regime.

At a rotation frequency of $\rotfreq=2\pi \times \SI{5}{\giga\hertz}$, the calculated permittivity is $\relpermitttensbcomp^{\xpos\xpos}=\relpermitttensbcomp^{\ypos\ypos}\simeq\num{2.04}$ in the equatorial plane and $\relpermitttensbcomp^{\zpos\zpos}=\num{1.98}$ along the rotation axis. The polarizability is increased by about $\SI{4.9}{\percent}$ in the equatorial plane and reduced by $\SI{2.3}{\percent}$ in the axial direction. Since the trap frequencies of nanoparticles levitated with optical tweezers scale with the square root of the polarizability~\cite{chang_cavity_2010,romero-isart_optically_2011}, these values correspond to an increase in the trap frequency along the $\xpos$ and $\ypos$ directions by about $\SI{2.5}{\percent}$ and a decrease of the trap frequency along the $\zpos$ direction by about $\SI{1.1}{\percent}$. These changes are in the $\si{\kilo\hertz}$ range in typical setups and large enough to be detected~\cite{meyer_resolved-sideband_2019,windey_cavity-based_2019,tebbenjohanns_cold_2019,delic_cavity_2019}.

\section{Conclusion}
\label{sec: conclusions}

To summarize, we provide a general theory of the interaction between the rotational degrees of freedom and the acoustic internal phonons of a nanoparticle within linear elastodynamics. We are able to model how the shape, the phonon spectrum, the permittivity, and the electric polarizability  of a nanoparticle are affected when it is spinning at a fixed frequency. By way of example, we explicitly calculate the dependence of these properties on the rotation frequency in the particular case of a dielectric nanosphere and show that its effects should be measurable at rotational frequencies recently  achieved experimentally.

The theory and results presented in this paper can be generalized to include anisotropic and inhomogeneous nanoparticles in orientation-dependent potentials, which is useful for other shapes such as nanorods~\cite{kuhn_optically_2017} or nanoplatelets~\cite{nagornykh_optical_2017}. Beyond the purely classical effects discussed here, the quantum theory employed in the paper can also be extended to address genuine quantum effects such as the phonon-induced decoherence of elastic rotors in macroscopic quantum superpositions~\cite{stickler_probing_2018,stickler_rotational_2018}. While we characterize at which rotation frequencies nonlinear elastic effects become relevant, our study is restricted to the linear elastic regime. Accounting for anharmonic corrections to the atom-atom interaction beyond the linear elastic approximation is an interesting future research direction which could for instance unveil tunable phonon-phonon interactions and provide richer phonon dynamics~\cite{juraschek_cavity_2019}. Further research directions (some of which we currently investigate) include: (i)\ using whispering gallery modes or evanescent coupling to photonic structures to measure changes in the nanoparticle geometry~\cite{childress_cavity_2017,aiello_perturbation_2019,magrini_near-field_2018}; (ii)\ studying Brillouin scattering off a levitated rotating nanoparticle~\cite{kuok_brillouin_2003};
(iii)\ studying the complex coupled dynamics between rotation, translation, and vibrations caused by the rotation-dependent polarizability and birefringence which modify the optical potential; (iv)\ studying more complex internal degrees of freedom such as spin waves (magnons) or electrons~\cite{gonzalez-ballestero_quantum_2020}; and (v)\ understanding the origin of the linewidths of such excitations~\cite{maccabe_phononic_2019} as well as their dependence on rotation-induced strain. We hope that this work will stimulate experiments exploring the internal mesoscopic quantum physics of levitated nanoparticles.

\begin{acknowledgments}
We thank the levitodynamics team in the group of L.~Novotny at ETH Zürich for inspiring discussions. D.~H. acknowledges support by the Studienstiftung des Deutschen Volkes. C.~G.~B.\ acknowledges funding from the EU Horizon 2020 program under the Marie Sk\l{}odowska-Curie grant agreement no.\ 796725.
\end{acknowledgments}

\appendix
\section{Review of Elastodynamics}
\label{sec: elastodynamics appendix}

Linear elasticity theory describes small deformations of a three-dimensional elastic body from its equilibrium shape~\cite{achenbach_wave_1973,eringen_elastodynamics_1975,gurtin_linear_1984}. The displacement field $\ufield(\post)$ indicates how far and in which direction each point $\pos$ of the body is displaced at a time $\tm$. The elastic properties of the body are described by the mass density $\dens(\pos)$ and the elasticity tensor $\elastens(\pos)$. The elasticity tensor is of fourth order, with symmetries $\elastenscomp^{ijkl} = \elastenscomp^{jikl} = \elastenscomp^{ijlk} = \elastenscomp^{klij}$~\cite{gurtin_linear_1984}. In case of a homogeneous and isotropic elastic body, $\dens$ and $\elastens$ are constant in space, and the latter has only two independent coefficients~\cite{gurtin_linear_1984,achenbach_wave_1973}:
\begin{equation}
  \elastenscomp^{ijkl} = \lamemu \spare{\kronecker^{ik}\kronecker^{jl} + \kronecker^{il}\kronecker^{jk}} + \lamelambda \kronecker^{ij}\kronecker^{kl}.
\end{equation}
Here, the two real-valued constants $\lamelambda$ and $\lamemu$ are called Lamé parameters. The elastic properties of a homogeneous and isotropic body are thus described by three numbers: $\dens$, $\lamelambda$, and $\lamemu$. Excitations of the displacement field can be both transverse and longitudinal and there are two distinct sound speeds for transverse and longitudinal waves,
\begin{align}\label{eqn: definition sound speeds}
  \ctrans &= \sqrt{\frac{\lamemu}{\dens}},  & \clong = & \sqrt{\frac{2\lamemu+\lamelambda}{\dens}},
\end{align}
respectively~\cite{achenbach_wave_1973}. In consequence, $\lamemu$ needs to be positive while $\lamelambda \geq -2\lamemu$. The equation of motion of the freely evolving displacement is
\begin{equation}\label{eqn: phonon equation of motion}
  \dens \ddeltt\ufield = \Dphon \ufield,
\end{equation}
where $\Dphon$ is a second-order differential operator acting as \mbox{$\spare{\Dphon\ufield }^i =\elastenscomp^{ijkl} \partial_j  \,\partial_k \ufieldcomp_l$}. A suitable set of boundary conditions is required to obtain a unique solution of the equation of motion given initial conditions. In case of a freely vibrating body, the boundary conditions are of Neumann type and state that the surface of the body is force free: $\stresstenscomp^{ij} \normveccomp^j = 0$ on the surface~\cite{gurtin_linear_1984}. Here, $\normveccomp^i$ are the components of the exterior surface normal vector field and $\stresstens$ is the stress tensor. The stress tensor describes the forces required to cause a deformation given by the strain tensor $\straintens$:
\begin{equation}\label{eqn: definition strain and stress tensor}
  \begin{split}
    \straintenscomp^{ij} &\equiv \frac{1}{2}\pare{\partial_i \ufieldcomp^j + \partial_j \ufieldcomp^i}, \\
    \stresstenscomp^{ij} &\equiv \elastenscomp^{ijkl}\straintenscomp^{kl}.
  \end{split}
\end{equation}

The vibrational eigenmodes $\wmode_\phonindex(\pos)$ of a linear elastic body and their spectrum of frequencies $\phonfreq_\phonindex$ are obtained as solutions of the eigenvalue equation
\begin{equation}\label{eqn: phonon eigenvalue equation}
  \Dphon \wmode_\phonindex(\pos) = -\dens \phonfreq^2_\phonindex \wmode_\phonindex(\pos)
\end{equation}
with the appropriate boundary conditions. Different eigenmodes are orthogonal and we take them to be normalized according to
\begin{equation}\label{eqn: phonon orthonormality condition}
  \int_\body \cconj\wmode_\phonindex(\pos) \cdot \wmode_\phonindexb(\pos)\,\dd\pos = \kronecker_{\phonindex \phonindexb}.
\end{equation}
Since $(-\Dphon)$ with the boundary conditions of a force-free body is a self-adjoint operator~\cite{anghel_quantization_2007}, the frequencies $\phonfreq_\phonindex$ are real-valued and the eigenmode solutions form a basis for the space of displacement fields. Any solution to the equation of motion \cref{eqn: phonon equation of motion} can then be expressed as a linear combination of eigenmodes, \mbox{$\ufield(\post) = \sum_\phonindex \ufieldmodedens_\phonindex \spare{\phonnormvar_\phonindex e^{-\im \phonfreq_\phonindex \tm}\wmode_\phonindex(\pos) + \cc}$}, where the normal variables $\phonnormvar_\phonindex \in \C$ are determined by the initial conditions and we define \mbox{$\ufieldmodedens_\phonindex \equiv \sqrt{\hbar/2\dens\phonfreq_\phonindex}$}.

Canonical quantization in terms of the eigenmodes can now be performed in the usual manner~\cite{cohen-tannoudji_photons_2004} starting from the Hamiltonian density
\begin{equation}\label{eqn: elastodynamics standard Hamiltonian}
  \Hamildens_0 \equiv \frac{\pifield^2}{2\dens} - \frac{1}{2} \ufield\cdot \Dphon\ufield,
\end{equation}
of a free elastic body obtained from the standard Lagrangian introduced in \cref{eqn: lagrangian elastic rotor}, where $\pifield \equiv \dens \ddelt \ufield$ is the conjugate momentum of the displacement. Quantization amounts to replacing the normal variables $\phonnormvar_\phonindex$ with ladder operators $\phonaop_\phonindex$ that satisfy the canonical commutation relations $[\phonaop_\phonindex,\hconj\phonaop_\phonindexb]=\kronecker_{\phonindex\phonindexb}$ due to the normalization \cref{eqn: phonon orthonormality condition}. Hence, the displacement field operator is
\begin{equation}\label{eqn: phonon field operator mode expansion}
    \ufieldop(\pos) = \sum_\phonindex \ufieldmodedens_\phonindex \spare{ \phonaop_\phonindex \wmode_\phonindex(\pos)  + \hc},
\end{equation}
and the resulting quantum Hamiltonian
\begin{equation}\label{eqn: bare phonon Hamiltonian}
  \HamilopBare = \hbar \sum_{\phonindex} \phonfreq_\phonindex \hconj\phonaop_\phonindex \phonaop_\phonindex.
\end{equation}

\section{Linear Elastic Rotor}
\label{sec: linear elastic rotor appendix}

In this appendix, we justify the Lagrangian \cref{eqn: lagrangian elastic rotor} serving as the cornerstone of this work. It describes the joint dynamical evolution of the rotation and vibration of an elastic body and can be obtained from a standard microscopic model of an elastic solid~\cite{ashcroft_solid_1976}. To this end, we consider the body as a system of a finite number of constituent point masses $\mass_\massindex$ (atoms) with positions $\poslab_\massindex$ relative to an inertial Cartesian laboratory frame $\labframe \equiv \{\pos_0;\exvec,\eyvec,\ezvec\}$ that has an arbitrary origin $\pos_0$ and an orientation determined by the orthonormal basis $\{\exvec,\eyvec,\ezvec\}$. The masses interact pairwise through a common potential $\potinternal$ which only depends on the relative position of each pair of masses. A Lagrangian describing the dynamics of such a system is
\begin{equation}\label{eqn: microscopic lagrangian}
  \Lagrangian = \frac{1}{2} \sum_\massindex \mass_\massindex (\velocitylab_\massindex)^2 - \frac{1}{2} \sum_{\massindex,\massindexb\neq \massindex} \potinternal(\poslab_\massindex-\poslab_\massindexb).
\end{equation}
A series of changes of variables allows us to describe the mechanics of the body in terms of its center of mass position $\CMpos$, its overall orientation $\eulerangles$, and the displacement of each mass from its equilibrium position within the body. The dynamics of these displacements can in turn be modeled by the displacement field $\ufield(\post)$ of linear elasticity theory by making suitable approximations.

The first step is to describe the position of the constituent masses relative to the center of mass position $\CMpos \equiv \sum_\massindex \mass_\massindex \pos_\massindex/\bodymass$ where $\bodymass \equiv \sum_\massindex \mass_\massindex$ is the total mass of the body:
\begin{equation}\label{eqn: comoving reference frame}
  \poscomov_\massindex \equiv \poslab_\massindex - \CMpos .
\end{equation}
This change of variables corresponds to introducing a comoving reference frame $\{\CMpos;\exvec,\eyvec,\ezvec\}$ which originates at $\CMpos$ and moves with velocity $\CMvelocity$ with respect to the laboratory frame. The defining property of a comoving reference frame is that the total linear momentum vanishes relative to such a frame, that is, $\sum_\massindex \mass_\massindex \ddelt\poscomov_\massindex = 0$.

Next, we describe the relative positions $\poscomov_\massindex$ with respect to a body frame $\bodyframe\equiv\{\CMpos;\eAvec,\eBvec,\eCvec\}$ that has an orientation determined by the time-dependent orthonormal basis $\{\eAvec,\eBvec,\eCvec\}$ and rotates with respect to the laboratory frame with an angular velocity $\rotfreqvec$. The two reference frames are connected by a time-dependent rotation matrix $\rotmatrix$ such that the components $\poscomovcomp_\massindex^\eta \equiv \poscomov_\massindex \cdot \unitvec_\eta$ relative to the comoving frame are related to the components $\poscorotcomp_\massindex^i \equiv \poscomov_\massindex \cdot \unitvec_i$ relative to the body frame by
\cite{goldstein_classical_2002}
\begin{equation}\label{eqn: corotating reference frame}
  \poscorotcomp_\massindex^i = \rotmatrixcomp^{i\nu} \poscomovcomp_\massindex^\nu.
\end{equation}
In this appendix, we use Greek indices to indicate components with respect to $\{\exvec,\eyvec,\ezvec\}$ and Latin indices to indicate components with respect to $\{\eAvec,\eBvec,\eCvec\}$. Repeated indices are summed over. The rotation matrix can be parametrized in the $zyz$ convention~\cite{goldstein_classical_2002} as $\rotmatrix(\eulerangles) \equiv \rotmatrix_\zpos(\eulerC) \rotmatrix_\ypos(\eulerB) \rotmatrix_\zpos(\eulerA)$ using three Euler angles $\eulerangles \equiv \transp{(\eulerA, \eulerB, \eulerC)}$ and
\begin{align}
  \rotmatrix_\ypos(a) & \equiv \begin{pmatrix} \cos a & 0 & - \sin a \\ 0 & 1 & 0 \\ \sin a & 0 & \cos a \end{pmatrix}, \\
  \rotmatrix_\zpos(a) & \equiv \begin{pmatrix} \cos a & \sin a & 0 \\ -\sin a & \cos a & 0 \\ 0 & 0 & 1 \end{pmatrix}.
\end{align}
The rotation frequency of the body frame is related to the Euler angles through
\begin{equation}\label{eqn: angular velocity body frame in lab frame}
  \rotfreqvec = \ddelt\eulerA \eEulerA + \ddelt\eulerB \eEulerB + \ddelt\eulerC \eEulerC,
\end{equation}
where $\eEulerA$, $\eEulerB$, and $\eEulerC$ are the unit vectors along the time-dependent rotation axes with respect to which the Euler angles are defined. The rotation axes are~\cite{goldstein_classical_2002}
\begin{equation}\label{eqn: rotation axes in lab frame}
  \begin{split}
    \eEulerA &= \ezvec, \\
    \eEulerB &= - \sin \eulerA \exvec + \cos \eulerA \eyvec, \\
    \eEulerC &= \cos \eulerA \sin \eulerB \exvec + \sin \eulerA \sin\eulerB \eyvec + \cos\eulerB \ezvec .
  \end{split}
\end{equation}
In order to express the Lagrangian in terms of these new variables, we use that rotations preserve inner products, $\poscomovcomp_\massindex^\eta \poscomovcomp^\eta_\massindexb = \poscomovcomp_\massindex^i \poscomovcomp^i_\massindexb$ and that the inverse (transpose) of the rotation matrix is given by $\rotmatrixcomp^{i \eta}(\eulerA,\eulerB,\eulerC) = \rotmatrixcomp^{\eta i}(-\eulerC,-\eulerB,-\eulerA)$. Furthermore, the time derivative of the rotation matrix can be expressed in terms of the rotation frequency $\rotfreqvec$ as
\begin{equation}
  {\ddelt\rotmatrixcomp}^{i\eta} \poscomovcomp_\massindex^\eta = - \rotmatrixcomp^{i\eta} \LeviCivita^{\eta\nu\mu} \rotfreq^\nu \poscomovcomp_\massindex^\mu
\end{equation}
where $\LeviCivita$ is the Levi-Civita symbol. In order for $\rotfreqvec$ to represent the rotation frequency of the elastic body as a whole, we need to choose it such that the total angular momentum vanishes relative to the body frame $\bodyframe$:
\begin{equation}\label{eqn: condition corotating frame}
  \sum_\massindex \mass_\massindex \LeviCivita^{ijk} \poscorotcomp^{j}_\massindex \ddelt\poscorotcomp^{k}_\massindex = 0.
\end{equation}
The body frame is then both a comoving and corotating reference frame.

Finally, we describe vibrations as displacements $\ufield_\massindex$ of each mass from its equilibrium position $\posequi_\massindex$ (i.e., the position minimizing the potential energy) relative to the body frame $\bodyframe$:
\begin{equation}\label{eqn: displacement}
  \ufield_\massindex \equiv \poscorot_\massindex - \posequi_\massindex .
\end{equation}
After applying the changes of variables \cref{eqn: comoving reference frame,eqn: corotating reference frame,eqn: displacement}, the Lagrangian \cref{eqn: microscopic lagrangian} takes the form
\begin{multline}\label{eqn: lagrangian of discrete elastic rotor}
  \Lagrangian = \frac{1}{2}\bodymass \CMvelocity^2 + \frac{1}{2} \rotfreq^i \inertialtenscomp^{ij} \rotfreq^j \\
  + \sum_\massindex \mass_\massindex \ddelt\ufield_\massindex^2 - \frac{1}{2} \sum_{\massindex,\massindexb\neq \massindex} \potinternal( \posequi_\massindex + \ufield_\massindex- \posequi_\massindexb - \ufield_\massindexb ).
\end{multline}
Here, $\inertialtens$ is the time-dependent inertial tensor of the body with components
\begin{equation}
  \inertialtenscomp^{ij} \equiv \sum_\massindex \mass_\massindex \spare{ \pare{ \posequicomp^i_\massindex + \ufieldcomp^i_\massindex }^2 \kronecker^{ij} - \pare{  \posequicomp^i_\massindex + \ufieldcomp^i_\massindex } \pare{ \posequicomp^j_\massindex + \ufieldcomp^j_\massindex } },
\end{equation}
relative to the body frame $\bodyframe$. The inertial tensor accounts for the actual mass distribution of the elastic body modified by vibrations around its equilibrium shape.

We can now pass from a point-mass model to a continuum description of the body in the usual manner~\cite{ashcroft_solid_1976}. This transition hinges on several approximations: (i)\ The displacements are considered to be small, which allows us to approximate the interaction potential $\potinternal$ as harmonic around the equilibrium shape of the body; (ii)\ The wavelength of elastic waves is assumed to be large compared to the distance of the individual masses forming the solid such that neighboring masses are subject to almost the same displacement; (iii)\ We take the continuum limit, which amounts to replacing the point mass distribution with a mass density distribution $\dens(\pos)$ and the individual displacements $\ufield_\massindex$ with the continuous displacement field $\ufield(\post)$. By extending this standard procedure to the case of a rotating body \cref{eqn: lagrangian of discrete elastic rotor} considered here, we obtain a Lagrangian generating the translational, rotational, and vibrational dynamics of a free linear elastic rotor:
\begin{multline}\label{eqn: lagrangian of elastic rotor}
    \Lagrangian = \frac{\bodymass}{2} \CMvelocity^2  + \int_\body \spare{\frac{\dens}{2}  \ddelt\ufield^2 - \frac{1}{2} \straintenscomp^{ij} \elastenscomp^{ijkl} \straintenscomp^{kl} } \dd \pos \\
    + \frac{1}{2} \rotfreq^i \inertialtenscomp^{ij}[\ufield] \rotfreq^j.
\end{multline}
The dynamical variables are now the center of mass $\CMpos(\tm)$, the orientation $\eulerangles(\tm)$ [appearing in the angular velocity $\rotfreqvec$], and the displacement field $\ufield(\post)$.
The first term in \cref{eqn: lagrangian of elastic rotor} represents the kinetic energy of the center of mass. The second term is the standard Lagrangian of linear elasticity theory~\cite{achenbach_wave_1973,eringen_elastodynamics_1975}. However, the displacement field $\ufield(\post)$ now describes vibrations relative to the comoving and corotating body frame $\bodyframe$ in contrast to the linear elasticity theory of resting bodies. The third term is the kinetic energy of the rotation of the body around its own axis. The inertial tensor is a functional of the displacement field:
\begin{multline}\label{eqn: inertial tensor functional}
  \inertialtenscomp^{ij}[\ufield] \equiv \int_\body \dens \big[ (\poscomp^k + \ufieldcomp^k) (\poscomp^k + \ufieldcomp^k)\kronecker^{ij} \\
  - (\poscomp^i + \ufieldcomp^i)(\poscomp^j + \ufieldcomp^j) \big] \dd \pos.
\end{multline}
The inertial tensor is symmetric, $\inertialtenscomp^{ij} = \inertialtenscomp^{ji}$ and the case of a rigid rotor is recovered by setting $\ufield = 0$. The equations of motion of the displacement field and the Euler angles $a \in \{\eulerA,\eulerB,\eulerC\}$
\begin{equation}\label{eqn: eom displacement field and euler angles}
  \begin{split}
    \dens \ddeltt\ufieldcomp^i = [\Dphon \ufield]^i + \dens \spare{ (\poscomp^i + \ufieldcomp^i) \rotfreq^j\rotfreq^j - ( \poscomp^j + \ufieldcomp^j )\rotfreq^i \rotfreq^j },\\
    \ddelt\unitveccomp_{a}^i  \inertialtenscomp^{ij}[\ufield] \rotfreq^j + \unitveccomp^i_{a}  \inertialtenscomp^{ij}[\ufield] \ddelt\rotfreq^j= \fpd{\rotfreq^i}{a}  \inertialtenscomp^{ij}[\ufield] \rotfreq^j - \unitveccomp^i_a \ddelt\inertialtenscomp^{ij}[\ufield] \rotfreq^j
  \end{split}
\end{equation}
are coupled while the center of mass decouples from the other degrees of freedom, $\CMvelocity=0$. The displacement field in particular is subject to centrifugal forces as discussed in \cref{sec: theory}; In contrast, it does not experience Euler or Coriolis forces due to the manner in which we define the corotating frame in \cref{eqn: condition corotating frame}. The familiar equation of motion of linear elastodynamics in the absence of any rotation is recovered in the case $\rotfreqvec=\zerovec$.

The Lagrangian \cref{eqn: lagrangian of elastic rotor} can easily be extended to describe nanoparticles which are levitated or whose orientation is externally controlled by subtracting an external potential $\potexternal(\CMpos,\eulerangles)$ that is quadratic and hence does not couple the translational to rotational and vibrational degrees of freedom. Moreover, one can perform canonical quantization of the corresponding Hamiltonian  in order to obtain a full quantum description of the translational, rotational, and vibrational degrees of freedom of a linear elastic rotor~\cite{rusconi_magnetic_2016,stickler_spatio-orientational_2016,stickler_rotational_2018}.

In the body of this paper, we study the effect rotation has on vibrations of an elastic body in a regime where the modulation of the rotational frequency $\rotfreqvec$ of the rotor due to vibrations can be neglected. To this end, we assume that the rotation frequency is constant and that the body is levitated in a quadratic potential $\potexternal(\CMpos)$. Hence, the displacement field is the only remaining dynamical variable. The classical Hamilton functional resulting from \cref{eqn: lagrangian of elastic rotor} in this case is of the form
\begin{equation}\label{eqn: Hamilton functional fixed rotation}
    \Hamiltonian = \int_\body \spare{ \Hamildens_0 + \Hamildens_1 + \Hamildens_2 } \dd\pos.
\end{equation}
The first term is the Hamiltonian of linear elastodynamics \cref{eqn: elastodynamics standard Hamiltonian}. The second and third terms derive from the corrections to the inertial tensor \cref{eqn: inertial tensor functional} and are of first and second order in the displacement field, respectively:
\begin{equation}
  \begin{split}
    \Hamildens_1 &\equiv - \frac{\dens}{2} \rotfreq^i \spare{ 2 \ufieldcomp^{k}\poscomp^k \kronecker^{ij} - ( \ufieldcomp^i \poscomp^j + \poscomp^i \ufieldcomp^j ) } \rotfreq^j,\\
     \Hamildens_2 &\equiv - \frac{\dens}{2} \rotfreq^i \spare{ \ufieldcomp^k\ufieldcomp^k  \kronecker^{ij} - \ufieldcomp^i  \ufieldcomp^j } \rotfreq^j.
  \end{split}
\end{equation}
They describe the centrifugal forces acting on the displacement field \footnote{Note that these terms would also be present in the full Hamiltonian corresponding to \cref{eqn: lagrangian of elastic rotor}, but with opposite sign.}. Canonical quantization of this Hamiltonian leads to the quantum Hamiltonian \cref{eqn: fixed rotation total Hamiltonian} used throughout this paper.

\section{Elastic Sphere}
\label{sec: sphere appendix}

In this appendix, we summarize results for the particular case of a spherical particle. In \cref{sec: sphere eigenmodes appendix}, we revise the known mode structure of a resting linear elastic sphere. In \cref{sec: spinning sphere appendix}, we list our results for a spinning elastic sphere that we use in the case study in \cref{sec: case study}.

\subsection{Eigenmodes of a Resting Sphere}
\label{sec: sphere eigenmodes appendix}

The elastic eigenmodes and frequency spectrum of a homogeneous and isotropic sphere of radius $\rad$ centered at the origin of the coordinate system can be obtained by solving \cref{eqn: phonon eigenvalue equation} with the appropriate boundary conditions and are well known~\cite{eringen_elastodynamics_1975,achenbach_wave_1973}. We summarize the relevant results in a form suitable for this paper, using spherical coordinates $(\rpos,\polpos,\azpos)$ such that $\poscomp^1 = \rpos \sin\polpos \cos\azpos$, $\poscomp^2 = \rpos \sin\polpos \sin\azpos$, and $\poscomp^3 = \rpos \cos\polpos$. Here, $\poscomp^i$ are the Cartesian components of the position vector in the body frame; see \cref{sec: linear elastic rotor appendix}. All solutions of the eigenmode equation \cref{eqn: phonon eigenvalue equation} for a given phonon frequency $\phonfreq$ can conveniently be expressed in spherical coordinates and in terms of a set of vector spherical harmonics~\cite{barrera_vector_1985}
\begin{equation}
  \begin{split}
    \VSHY{l}{m}(\polpos,\azpos) &\equiv \ervec \sphericalY{l}{m}(\polpos,\azpos),\\
    \VSHPsi{l}{m}(\polpos,\azpos) &\equiv \rpos \grad \sphericalY{l}{m}(\polpos,\azpos),\\
    \VSHPhi{l}{m}(\polpos,\azpos) &\equiv \pos \times \grad \sphericalY{l}{m}(\polpos,\azpos),
  \end{split}
\end{equation}
where we follow the convention of~\cite{jackson_classical_1999} in the definition of the spherical harmonics $\sphericalY{l}{m}$. The vector field $\VSHY{l}{m}$ is purely radial while $\VSHPsi{l}{m}$ and $\VSHPhi{l}{m}$ have only polar and azimuthal components. The space of solutions that are finite in the volume of the sphere is spanned by displacement modal fields $\wmode_\phonindex(\pos)$ that are of the from given in \cref{tab: phonon displacement modal fields}, with a mode index $\phonindex$ containing the polar order \mbox{$\phonl \in \N_0$} and the azimuthal order \mbox{$\phonm \in \Z$}, \mbox{$|\phonm|\leq \phonl$}. We express the modal fields in terms of dimensionless quantities defined in \cref{tab: phonon quantities}. The radial constants $\phona$ and $\phonb$ in particular determine how rapidly the displacement modal field oscillates in the radial direction.

\begin{table}
  \setlength{\extrarowheight}{4pt}
  \newcolumntype{A}{>{\begin{math}}r<{\end{math}}}
  \newcolumntype{B}{>{\begin{math}}c<{\end{math}}}
  \newcolumntype{C}{>{\begin{math}}l<{\end{math}}}
  \begin{tabularx}{\columnwidth}[t]{A B C }
    \toprule
    \wmode_\phonindex(\pos) &=& \rad^{-3/2} \big[ \wmodercomp^\VSHYcomp_{\phonl,\phonm}\pare{\rpos/\rad} \VSHY{\phonl}{\phonm}(\polpos,\azpos) \\
     &&+ \wmodercomp^\VSHPsicomp_{\phonl,\phonm}\pare{\rpos/\rad} \VSHPsi{\phonl}{\phonm}(\polpos,\azpos)
      + \wmodercomp^\VSHPhicomp_{\phonl,\phonm}\pare{\rpos/\rad} \VSHPhi{\phonl}{\phonm}(\polpos,\azpos) \big]\\
      \bottomrule
    \end{tabularx}
    \setlength{\extrarowheight}{4pt}
    \begin{tabularx}{\columnwidth}[t]{A @{${}={}$} C }
    \wmodercomp^\VSHYcomp_\phonindex(\rposR) & A_\phonindex \spare{\phonl \besselj{\phonl}(\phonaR\rposR)/\rposR - \phonaR \besselj{\phonl+1}(\phonaR \rposR)} + C_\phonindex [\phonl(\phonl+1)] \besselj{\phonl}(\phonbR \rposR)/\rposR\\
    \wmodercomp^\VSHPsicomp_\phonindex(\rposR) &  A_\phonindex \besselj{\phonl}(\phonaR \rposR)/\rposR + C_\phonindex \spare{ (\phonl+1) \besselj{\phonl}(\phonbR \rposR)/\rposR - \phonbR \besselj{\phonl+1}(\phonbR \rposR)  } \\
    \wmodercomp^\VSHPhicomp_\phonindex(\rposR) & B_\phonindex \besselj{\phonl}(\phonbR\rposR) \\
    \bottomrule
  \end{tabularx}
  \caption{Displacement modal field of the phonon eigenmodes of a sphere. The radial partial waves $\wmodercomp^\VSHYcomp$, $\wmodercomp^\VSHPsicomp$, and $\wmodercomp^\VSHPhicomp$ contain spherical Bessel functions $\besselj{\phonl}$ of the first kind and have amplitudes $A_\phonindex$, $B_\phonindex$, and $C_\phonindex$. The amplitudes are determined by the boundary conditions and listed explicitly in \cref{tab: phonon displacement modal field amplitudes}. All other parameters are defined in \cref{tab: phonon quantities}.}
  \label{tab: phonon displacement modal fields}
\end{table}

\begin{table}
  \newcolumntype{A}{>{\begin{math}}r<{\end{math}}}
  \newcolumntype{C}{>{\begin{math}}X<{\end{math}}}
  \begin{tabularx}{\columnwidth}[t]{A @{${}={}$} C @{\qquad\qquad} A @{${}={}$} C  @{\quad}  A @{${}={}$} C}
    \toprule
    \clong & \sqrt{(2\lamemu+\lamelambda)/\dens} & \ctrans & \sqrt{\lamemu/\dens} \\
    \phona & \phonfreq/\clong & \phonb & \phonfreq/\ctrans \\
    \phonaR & \phona\rad & \phonbR & \phonb\rad & \rposR & \rpos/ \rad\\
    \bottomrule
  \end{tabularx}
  \caption{Definitions of the longitudinal and the transverse sound velocity $\clong$ and $\ctrans$, as well as the radial constants $\phona$ and $\phonb$, and the dimensionless quantities appearing in the phonon modal fields. The definitions are given in terms of the density $\dens$, Lamé constants $\lamelambda,\lamemu$, fiber radius $\rad$, and radial position $\rpos$.}
  \label{tab: phonon quantities}
\end{table}

\begin{table}
  \setlength{\extrarowheight}{4pt}
  \newcolumntype{A}{>{\begin{math}}r<{\end{math}}}
  \newcolumntype{B}{>{\begin{math}}c<{\end{math}}}
  \newcolumntype{C}{>{\begin{math}\thinmuskip=.7mu\medmuskip=1mu}l<{\end{math}}}
  \begin{tabularx}{\columnwidth}[t]{A B C}
    \toprule
    \stressmodecomp^{\rpos\rpos}_\phonindex(\pos) &=& \lamemu/(\rpos^2\sqrt{\rad}) [ A_\phonindex M_{11}\pare{\rpos/\rad} - \im B_\phonindex M_{12}\pare{\rpos/\rad} \\
    && + C_\phonindex M_{13}\pare{\rpos/\rad} ]  \legendreP{\phonl}{\phonm}(\cos\polpos) e^{\im \phonm \azpos}\\
    \stressmodecomp^{\rpos\polpos}_\phonindex(\pos) &=& \lamemu/(\rpos^2\sqrt{\rad})[ A_\phonindex M_{21}\pare{\rpos/\rad} \del_\polpos - \im B_\phonindex M_{22}\pare{\rpos/\rad} \phonm/\sin\polpos \\
    && + C_\phonindex M_{23}\pare{\rpos/\rad}\del_\polpos ]  \legendreP{\phonl}{\phonm}(\cos\polpos) e^{\im \phonm \azpos} \\
    \stressmodecomp^{\rpos\azpos}_\phonindex(\pos) &=& \im \lamemu/(\rpos^2\sqrt{\rad}) [ A_\phonindex M_{31}\pare{\rpos/\rad} \phonm/\sin\polpos - \im B_\phonindex M_{32}\pare{\rpos/\rad}\del_\polpos \\
    && + C_\phonindex M_{33}\pare{\rpos/\rad} \phonm/\sin\polpos ]  \legendreP{\phonl}{\phonm}(\cos\polpos) e^{\im \phonm \azpos} \\
    \bottomrule
  \end{tabularx}
  \setlength{\extrarowheight}{2pt}
  \newcolumntype{D}{>{\begin{math}}l<{\end{math}}}
  \begin{tabularx}{\columnwidth}[t]{A B D}
    M_{11}(\rposR) &=& \spare{2\phonl(\phonl-1)-\phonbR^2 \rposR^2} \besselj{\phonl}(\phonaR \rposR) + 4 \phonaR \rposR \besselj{\phonl+1}(\phonaR \rposR) \\
    M_{12}(\rposR) &=& 0 \\
    M_{13}(\rposR) &=& 2\phonl(\phonl+1) \spare{(\phonl-1)\besselj{\phonl}(\phonbR \rposR) - \phonbR \rposR \besselj{\phonl+1}(\phonbR \rposR)} \\
    M_{21}(\rposR) &=& 2(\phonl-1)\besselj{\phonl}(\phonaR \rposR) - 2\phonaR \rposR \besselj{\phonl+1}(\phonaR \rposR) \\
    M_{22}(\rposR) &=& (\phonl-1) \rposR \besselj{\phonl}(\phonbR \rposR) - \phonbR \rposR^2 \besselj{\phonl+1}(\phonbR \rposR) \\
    M_{23}(\rposR) &=& \spare{2(\phonl^2-1) - \phonbR^2 \rposR^2} \besselj{\phonl}(\phonbR \rposR) + 2\phonbR \rposR \besselj{\phonl+1}(\phonbR \rposR) \\
    M_{31}(\rposR) &=& M_{21}(\rposR) \\
    M_{32}(\rposR) &=& M_{22}(\rposR) \\
    M_{33}(\rposR) &=& M_{23}(\rposR) \\
    \bottomrule
  \end{tabularx}
  \caption{Radial components of the stress modal fields related to displacement modal fields in \cref{tab: phonon displacement modal fields} by \cref{eqn: definition strain and stress tensor}.}
  \label{tab: sphere phonon stress modal fields}
\end{table}

\begin{figure*}
  \includegraphics[height=164.51pt]{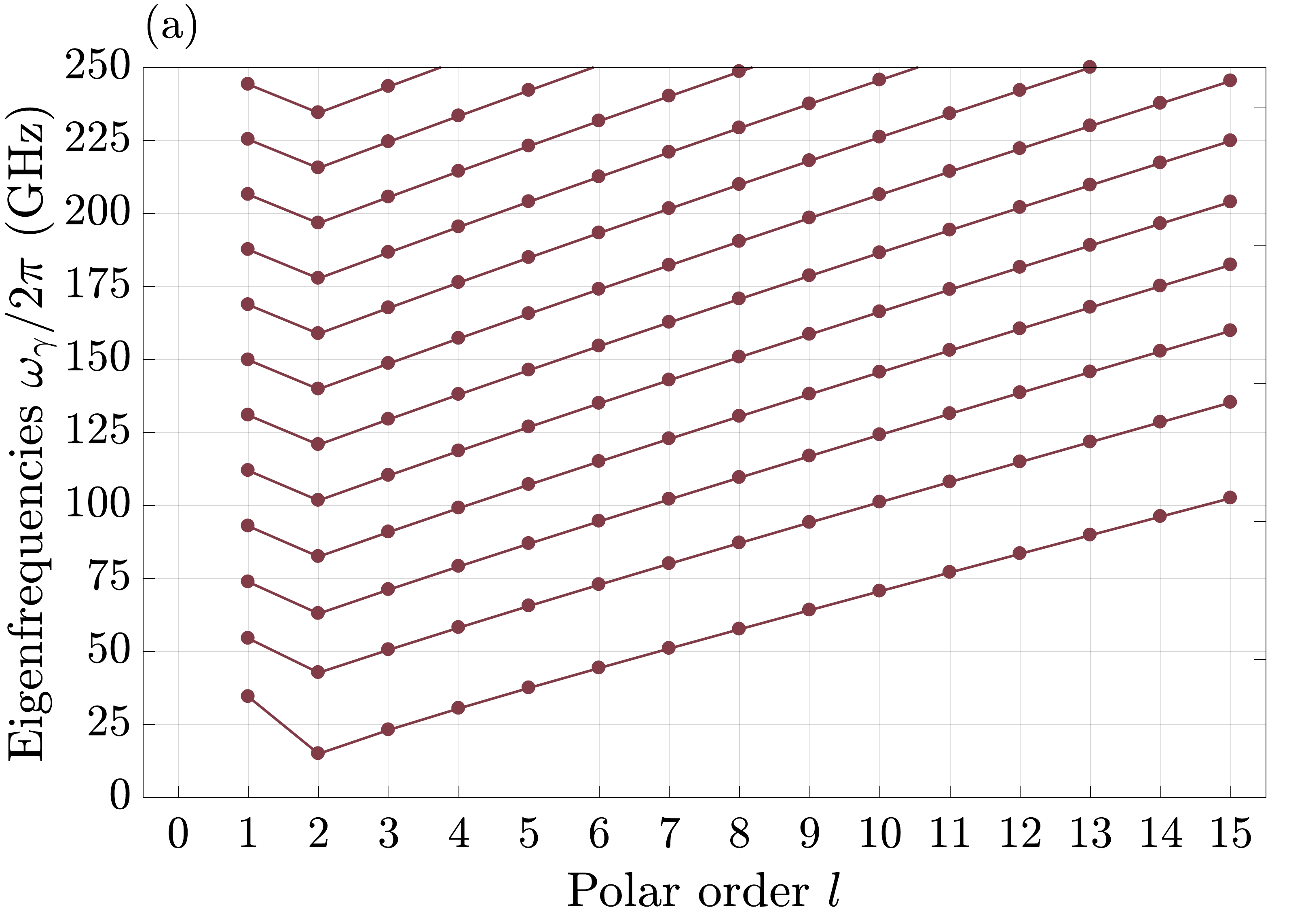}
  \includegraphics[height=164.51pt]{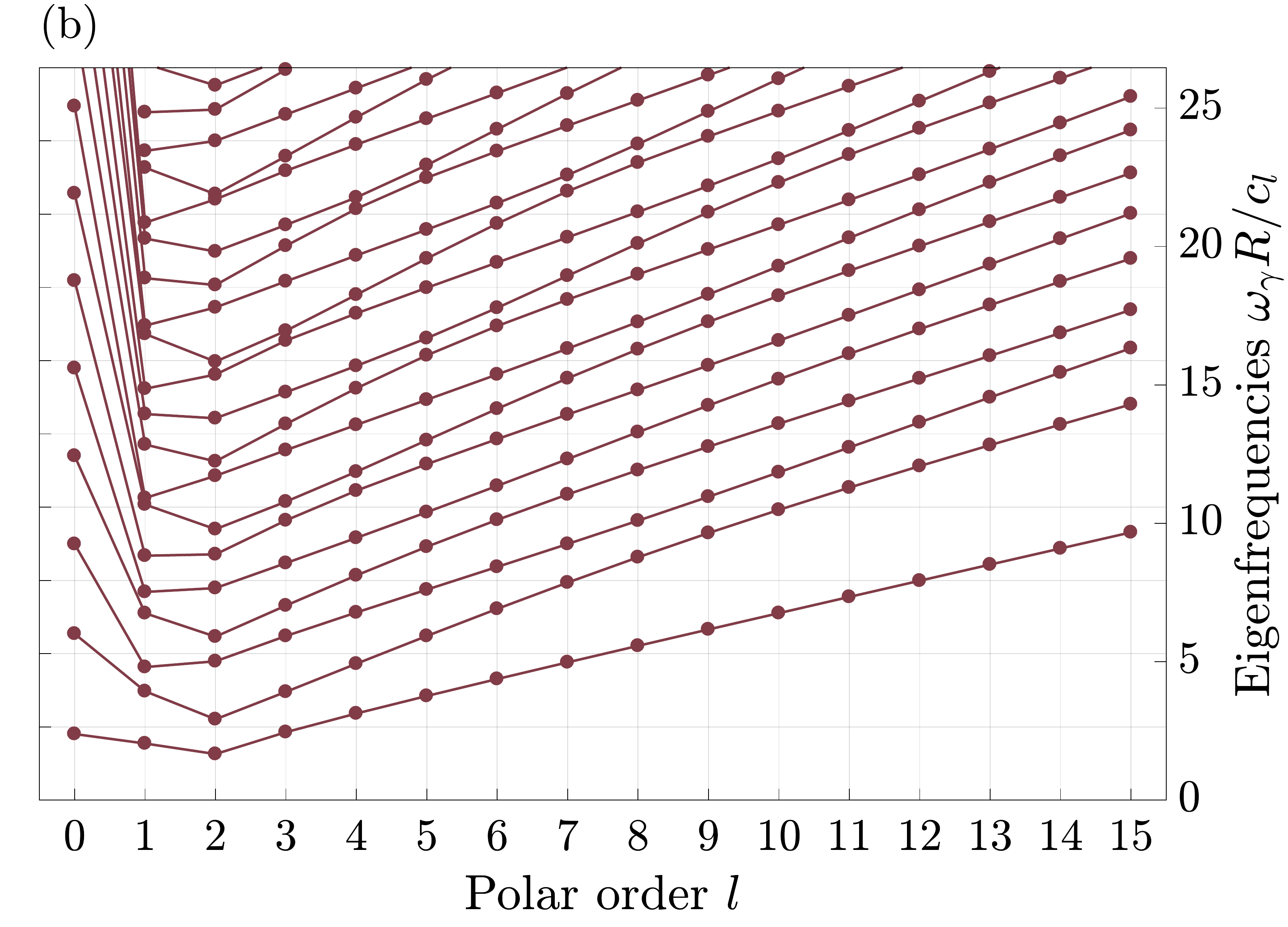}
  \caption{Frequency spectrum of the phonon eigenmodes of a resting nanosphere with properties specified in \cref{tab: parameters}. Panel~(a) shows the spectrum of the torsional modes as a function of the polar order $\phonl$, panel~(b) shows the spectrum of the spheroidal modes. The eigenfrequencies are indicated by dots while the lines serve as a guide to the eye. The order of magnitude of the lowest phonon frequency is determined by $\clong/\rad$. In units of this fundamental frequency, the eigenfrequencies of the torsional modes are independent of any system parameters and the eigenfrequencies of the spheroidal modes depend only on the Poisson ratio $\Poissonnu$.}
  \label{fig: sphere phonon spectrum}
\end{figure*}

\begin{table*}
\newcolumntype{D}{p{4cm}}
\newcolumntype{A}{>{\begin{math}}r<{\end{math}}}
\newcolumntype{B}{>{\begin{math}}c<{\end{math}}}
\newcolumntype{C}{>{\begin{math}}l<{\end{math}}}
\newcolumntype{E}{>{\begin{math}}X<{\end{math}}}
\begin{tabularx}{\textwidth}[t]{D ABC @{\qquad\qquad} ABC @{\qquad\qquad}ABE}
  \toprule
  Case & \multicolumn{9}{c}{Amplitudes}\\
  \midrule
  Torsional modes & A_\phonindex &=& 0 & B_\phonindex &=& \sqrt{2}/\sqrt{\phonl(\phonl+1) \spare{\besselj{l}^2(\phonbR_\phonindex) - \besselj{l-1}(\phonbR_\phonindex)\besselj{l+1}(\phonbR_\phonindex)}} & C_\phonindex &=& 0 \\
  Spheroidal modes,  $\phonl=0$ & A_\phonindex &=& 1/\sqrt{ \phonaR_\phonindex^2 I_2(1,\phonaR) } & B_\phonindex &=& 0 &  C_\phonindex &=& 0 \\
  Spheroidal modes, $\phonl\geq1$ & A_\phonindex &=& \multicolumn{7}{l}{$\Big[ \phonaR_\phonindex \besselj{\phonl}(\phonaR_\phonindex) \besselj{\phonl}'(\phonaR_\phonindex) + c_\phonindex^2\phonl(\phonl+1) \phonbR_\phonindex \besselj{\phonl}(\phonbR_\phonindex)\besselj{\phonl}'(\phonbR_\phonindex) + 2 c_\phonindex \phonl(\phonl+1) \besselj{\phonl}(\phonaR_\phonindex) \besselj{\phonl}(\phonbR_\phonindex) + \phonaR_\phonindex^2 I_2(\phonl,\phonaR_\phonindex)$} \\
   & & & \multicolumn{7}{r}{$+ c_\phonindex^2 \phonl(\phonl+1)(2\phonl+1) I_1(\phonl,\phonbR_\phonindex) + c_\phonindex^2 \phonl(\phonl+1) \phonbR_\phonindex^2 I_2(\phonl,\phonbR_\phonindex) - 2 c_\phonindex^2 \phonl(\phonl+1) \phonbR_\phonindex I_3(\phonl,\phonbR_\phonindex) \Big]^{-1/2}$} \\
   & B_\phonindex &=& 0 & C_\phonindex &=& c_\phonindex A_\phonindex \\
   \bottomrule
\end{tabularx}
\setlength{\extrarowheight}{5pt}
\begin{tabularx}{\textwidth}[t]{ABE}
  c_\phonindex & \equiv& - \spare{ 2 (\phonl-1) \besselj{\phonl}(\phonaR_\phonindex) - 2 \phonaR_\phonindex  \besselj{\phonl+1}(\phonaR_\phonindex) }/\spare{ \pare{2 \phonl^2-2-\phonbR_\phonindex^2 } \besselj{\phonl}(\phonbR_\phonindex )+2 \phonbR_\phonindex \besselj{\phonl+1}(\phonbR_\phonindex ) } \\
  I_1(\phonl,\phonaR) &\equiv& \int_0^1 \besselj{\phonl}^2(\phonaR \rposR)\, \dd \rposR
  = \pi \phonaR^{2 \phonl} \spare{(2 \phonl+1)4^{\phonl+1}
  \Gamma(\phonl+3/2)^2}^{-1} ~ \tensor[_2]{F}{_3}\pare{\phonl+1/2,\phonl+1; \, \phonl+3/2, \phonl+3/2, 2 \phonl+2; \, -\phonaR^2}\\
  I_2(\phonl,\phonaR) &\equiv& \int_0^1 \rposR^2 \besselj{\phonl}^2(\phonaR \rposR) \,\dd \rposR
  = \spare{ \besselj{\phonl}^2(\phonaR) + \besselj{\phonl+1}^2(\phonaR) - (2 \phonl+1) \besselj{\phonl}(\phonaR) \besselj{\phonl+1}(\phonaR)/\phonaR } /2\\
  I_3(\phonl,\phonaR) &\equiv& \int_0^1 \rposR \besselj{\phonl}(\phonaR \rposR) \besselj{\phonl+1}(\phonaR \rposR) \,\dd \rposR
  = \pi \phonaR^{2\phonl+1} \spare{4^{\phonl+2} \Gamma(\phonl+5/2)^2}^{-1} ~ \tensor[_2]{F}{_3} \pare{ \phonl+3/2,\phonl+2;\, \phonl+5/2, \phonl+5/2, 2\phonl+3; \,-\phonaR ^2}\\
    \bottomrule
\end{tabularx}
\caption{Amplitudes $A_\phonindex$, $B_\phonindex$, and $C_\phonindex$ of the displacement modal fields listed in \cref{tab: phonon displacement modal fields}. The functions $\besselj{\phonl}$ are spherical Bessel functions, $\tensor[_p]{F}{_q}$ are generalized hypergeometric functions, and $\Gamma$ is the gamma function; all other quantities are defined in \cref{tab: phonon quantities}. The amplitudes listed for the case of spheroidal modes with polar order $\phonl\geq1$ are valid in the generic case that all coefficients $M_{ij}$ in the frequency equation \cref{eqn: spheroidal mode frequency equation} are nonzero. If this is not the case, the amplitudes need to be recalculated using the boundary conditions \cref{eqn: boundary conditions matrix notation}.}
\label{tab: phonon displacement modal field amplitudes}
\end{table*}

\begin{table*}
\setlength{\extrarowheight}{4pt}
  \newcolumntype{A}{>{\begin{math}}r<{\end{math}}}
  \newcolumntype{B}{>{\begin{math}}c<{\end{math}}}
  \newcolumntype{C}{>{\begin{math}}l<{\end{math}}}
  \begin{tabularx}{\textwidth}{ABC}
    \toprule
    \strainmodecomp_\phonindex^{\rpos\rpos}(\pos) & = & A_\phonindex / (\rposR^2\sqrt{\rad^5}) \cpare{ \spare{ \phonl(\phonl-1) - \phonaR_\phonindex^2 \rposR^2 } \besselj{\phonl}(\phonaR_\phonindex \rposR) + 2 \phonaR_\phonindex \rposR \besselj{\phonl+1}(\phonaR_\phonindex\rposR) } \sphericalY{\phonl}{\phonm}(\polpos,\azpos)\\
    & & + \phonl(\phonl+1) C_\phonindex  / (\rposR^2\sqrt{\rad^5})\spare{ (\phonl-1) \besselj{\phonl}(\phonbR_\phonindex \rposR) - \phonbR_\phonindex \rposR \besselj{\phonl+1}(\phonbR_\phonindex \rposR) }\sphericalY{\phonl}{\phonm}(\polpos,\azpos) \\
    \strainmodecomp_\phonindex^{\rpos\polpos}(\pos) & = & A_\phonindex / (\rposR^2\sqrt{\rad^5}) \spare{(\phonl-1) \besselj{\phonl}(\phonaR_\phonindex\rposR) - \phonaR_\phonindex \rposR \besselj{\phonl+1}(\phonaR_\phonindex \rposR) } \del_\polpos \sphericalY{\phonl}{\phonm}(\polpos,\azpos) \\
    & & + C_\phonindex  / (\rposR^2\sqrt{\rad^5}) \cpare{ \spare{(\phonl^2-1) - \phonbR_\phonindex^2 \rposR^2 / 2 } \besselj{\phonl}(\phonbR_\phonindex\rposR) + \phonbR_\phonindex \rposR \besselj{\phonl+1}(\phonbR_\phonindex \rposR) } \del_\polpos \sphericalY{\phonl}{\phonm}(\polpos,\azpos)\\
    \strainmodecomp_\phonindex^{\rpos\azpos}(\pos) & = & \im \phonm A_\phonindex / (\rposR^2\sqrt{\rad^5}) \spare{ (\phonl-1) \besselj{\phonl}(\phonaR_\phonindex\rposR) - \phonaR_\phonindex \rposR \besselj{\phonl+1}(\phonaR_\phonindex\rposR) } \sphericalY{\phonl}{\phonm}(\polpos,\azpos)/\sin\polpos \\
    & & + \im \phonm C_\phonindex / (\rposR^2\sqrt{\rad^5}) \cpare{ \spare{ (\phonl^2-1) - \phonbR_\phonindex^2 \rposR^2/2 } \besselj{\phonl}(\phonbR_\phonindex \rposR) + \phonbR_\phonindex \rposR \besselj{\phonl+1}(\phonbR_\phonindex \rposR) } \sphericalY{\phonl}{\phonm}(\polpos,\azpos)/\sin\polpos \\
    \strainmodecomp_\phonindex^{\polpos\polpos}(\pos) & = & A_\phonindex / (\rposR^2\sqrt{\rad^5}) \spare{ \phonl \besselj{\phonl}(\phonaR_\phonindex \rposR) - \phonaR_\phonindex \rposR \besselj{\phonl+1}(\phonaR_\phonindex \rposR) } \sphericalY{\phonl}{\phonm}(\polpos,\azpos)
    + C_\phonindex / (\rposR^2\sqrt{\rad^5}) \phonl(\phonl+1) \besselj{\phonl}(\phonbR_\phonindex \rposR) \sphericalY{\phonl}{\phonm}(\polpos,\azpos) \\
    & & + A_\phonindex / (\rposR^2\sqrt{\rad^5}) \besselj{\phonl}(\phonaR_\phonindex\rposR) \del^2_\polpos \sphericalY{\phonl}{\phonm}(\polpos,\azpos)
    + C_\phonindex / (\rposR^2\sqrt{\rad^5}) \spare{ (\phonl+1) \besselj{\phonl}(\phonbR_\phonindex \rposR) - \phonbR_\phonindex \rposR \besselj{\phonl+1}(\phonbR_\phonindex \rposR) } \del^2_\polpos \sphericalY{\phonl}{\phonm}(\polpos,\azpos) \\
    \strainmodecomp_\phonindex^{\polpos\azpos}(\pos) & = & \im \phonm \cpare{ A_\phonindex / (\rposR^2\sqrt{\rad^5}) \besselj{\phonl}(\phonaR_\phonindex \rposR)+ C_\phonindex /  (\rposR^2\sqrt{\rad^5}) \spare{ (\phonl+1) \besselj{\phonl}(\phonbR_\phonindex \rposR) - \phonbR_\phonindex \rposR \besselj{\phonl+1}(\phonbR_\phonindex \rposR) } }  \spare{  \del_\polpos \sphericalY{\phonl}{\phonm}(\polpos,\azpos) -  \sphericalY{\phonl}{\phonm}(\polpos,\azpos)/\tan\polpos } / \sin\polpos\\
    \strainmodecomp_\phonindex^{\azpos\azpos}(\pos) & = & A_\phonindex / (\rposR^2\sqrt{\rad^5}) \spare{ \phonl \besselj{\phonl}(\phonaR_\phonindex \rposR) - \phonaR_\phonindex \rposR \besselj{\phonl+1}(\phonaR_\phonindex \rposR) } \sphericalY{\phonl}{\phonm}(\polpos,\azpos)
    +  C_\phonindex / (\rposR^2\sqrt{\rad^5}) \phonl(\phonl+1) \besselj{\phonl}(\phonbR_\phonindex\rposR) \sphericalY{\phonl}{\phonm}(\polpos,\azpos) \\ & &
    + \cpare { A_\phonindex / (\rposR^2\sqrt{\rad^5}) \besselj{\phonl}(\phonaR_\phonindex\rposR) + C_\phonindex / (\rposR^2\sqrt{\rad^5}) \spare{ (\phonl+1) \besselj{\phonl}(\phonbR_\phonindex \rposR) - \phonbR_\phonindex \rposR \besselj{\phonl+1}(\phonbR_\phonindex \rposR)  } } \spare{ \del_\polpos  \sphericalY{\phonl}{\phonm}(\polpos,\azpos)/\tan\polpos - \phonm^2 \sphericalY{\phonl}{\phonm}(\polpos,\azpos) / \sin\polpos }\\
    \bottomrule
  \end{tabularx}
  \caption{Strain modal fields of spheroidal modes related to the displacement modal fields in \cref{tab: phonon displacement modal fields} by \cref{eqn: definition strain and stress tensor}. The amplitudes $A_\phonindex$ and $C_\phonindex$ are specified in \cref{tab: phonon displacement modal field amplitudes} and the remaining quantities are defined in \cref{tab: phonon quantities}.}
  \label{eqn: strain modal fields spheroidal modes}
\end{table*}

In order to obtain the eigenmodes of a freely vibrating sphere, we need to impose as boundary condition that the stress in radial direction vanishes on the sphere surface, see \cref{sec: elastodynamics appendix}. The relevant components of the stress tensor modal field $\stressmode_\phonindex(\pos)$ corresponding to a displacement $\wmode_\phonindex(\pos)$ result from \cref{eqn: definition strain and stress tensor} and are given in \cref{tab: sphere phonon stress modal fields}. The boundary conditions are then equivalent to a set of linear equations for the amplitudes $A_\phonindex,  B_\phonindex, C_\phonindex$ of the displacement modal field (see \cref{tab: phonon displacement modal fields}):
\begin{align}\label{eqn: boundary conditions matrix notation}
  \begin{split}
    A_\phonindex M_{11} + C_\phonindex  M_{13} &= 0,\\
    A_\phonindex M_{21} + C_\phonindex  M_{23} &= 0,\\
    B_\phonindex M_{22} &=0
  \end{split}
\end{align}
where we abbreviate the coefficients $M_{ij} \equiv M_{ij}(1)$ of the stress modal fields defined in \cref{tab: sphere phonon stress modal fields}. The amplitudes $A_\phonindex$, $C_\phonindex$ on one hand and $B_\phonindex$ on the other are not coupled by the boundary conditions, resulting in two distinct mode families $\phonfam$: \emph{torsional modes} ($\phonfam=\Tmode$) and \emph{spheroidal modes} ($\phonfam=\Smode$).

\subsubsection{Torsional Modes}

Torsional eigenmodes are characterized by $A_\phonindex = C_\phonindex = 0$ and are purely transverse (that is, they feature a divergence-free displacement modal field). The displacement due to torsional modes is normal to the radial direction and changes neither the outward shape nor the density of the nanoparticle (to first order). The boundary conditions \cref{eqn: boundary conditions matrix notation} then require $M_{22} = 0$, that is, they impose the condition
\begin{equation}\label{eqn: torsional mode frequency equation}
  (\phonl-1) \besselj{\phonl}(\phonbR) - \phonbR \besselj{\phonl+1}(\phonbR) = 0
\end{equation}
on the dimensionless radial constant $\phonbR$ defined in \cref{tab: phonon quantities}. The roots $\phonbR_\phonindex = \phonfreq_\phonindex \rad/\ctrans$ of this frequency equation determine the discrete spectrum of frequencies $\phonfreq_\phonindex$ of the torsional eigenmodes, see panel~(a) in \cref{fig: sphere phonon spectrum}. For a given set $(\phonl,\phonm)$, we label the roots by $\phonn \in \N$ starting with $\phonn=1$ for the lowest frequency. We refer to $\phonn$ as the radial order since it counts the number of nodes of the modal field $\wmode(\pos)_\phonindex$ in $\rpos$ direction. There are no torsional modes for $\phonl=0$ and $\phonm=0$ because $\VSHPhi{0}{0}=0$. All torsional eigenmodes can therefore be labeled by mode indices $\phonindex = (\phonfam,\phonl,\phonm,\phonn)$ where
\begin{align}
  \phonfam&=\Tmode, & \phonl &\in \N, & \phonm &\in \Z, |\phonm| \leq \phonl, & \phonn &\in \N.
\end{align}
Each torsional mode can thus uniquely be identified by a term of the form $\Tmode_{\phonl\phonm\phonn}$.

The frequency equation \cref{eqn: torsional mode frequency equation} is independent of $\phonm$. In consequence, all modes $\Tmode_{\phonl\phonm\phonn}$, $\Tmode_{\phonl\phonmb\phonn}$ are degenerate in frequency. Moreover, the dimensionless roots $\phonbR_\phonindex$ of the frequency equation are universal in the sense that they are independent of the radius and the elastic properties of the sphere. The torsional eigenfrequencies hence scale with radius and sound speeds as
\begin{equation}\label{eqn: scaling phonon eigenfrequencies}
  \phonfreq_\phonindex = \frac{\ctrans}{\rad} \phonbR_\phonindex.
\end{equation}
The orthonormality condition \cref{eqn: phonon orthonormality condition} reduces to a normalization condition which determines the amplitude $B_\phonindex$ up to a complex phase. We choose the amplitude to be real valued, see \cref{tab: phonon displacement modal field amplitudes}. In this case, the displacement modal fields $\wmode_\phonindex$ obey the symmetry
\begin{equation}\label{eqn: displacement modal field azimuthal symmetry}
  \wmode_{-\phonm} = (-1)^\phonm \cconj\wmode_{\phonm}
\end{equation}
in the azimuthal order, dropping irrelevant mode indices. Moreover, the modal fields depend only on the eigenfrequency and the sphere radius, and scale with the latter as $\rad^{-3/2}$, see \cref{tab: phonon displacement modal fields}.

\begin{table*}
  \newcolumntype{A}{>{\begin{math}}r<{\end{math}}}
  \newcolumntype{B}{>{\begin{math}}c<{\end{math}}}
  \newcolumntype{C}{>{\begin{math}}l<{\end{math}}}
    \setlength{\extrarowheight}{2pt}
  \begin{tabularx}{\textwidth}{l@{\quad}ABC}
    \toprule
    Case & \multicolumn{3}{l}{Constant}\\
    \midrule
    \multicolumn{4}{l}{Linear shifts}\\
    $\phonfam = \Tmode$ &
    \linshift_\phonindex & = & 0\\
    $\phonfam = \Smode$ &
    \linshift_\phonindex & = & - \rotfreq^2\rad^3\sqrt{\dens/(\hbar\clong)} \, \sqrt{\phonaR_\phonindex}^{-1} \kronecker_{\phonm,0} \spare{ \kronecker_{\phonl,0} \sqrt{16\pi/18} \, I_{\phonindex}^Y - \kronecker_{\phonl,2} \sqrt{8\pi/45} (I_{\phonindex}^\VSHYcomp + 3 I_{\phonindex}^\VSHPsicomp) } \\
    \multicolumn{4}{l}{Beam splitter}\\
    $\phonfam = \Tmode, \phonfamb=\Tmode$ &
    \bscoupling_{\phonindex\phonindexb} & = & - (\rotfreq^2\rad/\clong) \phonaR_\phonindex^{-1} \kronecker_{\phonl\phonlb}\kronecker_{\phonm\phonmb}\kronecker_{\phonn\phonnb}  \spare{ 1-\phonm^2/(\phonl^2+\phonl) }/2  \\
    $\phonfam = \Smode, \phonfamb=\Smode$ &
    \bscoupling_{\phonindex\phonindexb} & = & - (\rotfreq^2\rad/\clong) \sqrt{\phonaR_\phonindex\phonaR_\phonindexb}^{-1} \kronecker_{\phonm\phonmb} \spare{ \kronecker_{\phonl\phonlb} \kronecker_{\phonn\phonnb}/2 + K_{-2} \,\kronecker_{\phonl-2,\phonlb}  + K_0 \,\kronecker_{\phonl\phonlb}  + K_{+2}\, \kronecker_{\phonl+2,\phonlb}  }\\
    $\phonfam = \Tmode, \phonfamb=\Smode$ &
    \bscoupling_{\phonindex\phonindexb} & = & - (\rotfreq^2\rad/\clong) \sqrt{\phonaR_\phonindex\phonaR_\phonindexb}^{-1} \kronecker_{\phonm\phonmb} \spare{ K_{-1} \,\kronecker_{\phonl-1,\phonlb}  + K_{+1} \,\kronecker_{\phonl+1,\phonlb} }\\
    $\phonfam = \Smode, \phonfamb=\Tmode$ &
    \bscoupling_{\phonindex\phonindexb} & = &
    \cconj\bscoupling_{\phonindexb\phonindex}\\
    \multicolumn{4}{l}{Two-mode squeezing}\\
    $\phonfam = \Tmode, \phonfamb=\Tmode$ &
    \sqcoupling_{\phonindex\phonindexb} & = & (-1)^{\phonm+1} (\rotfreq^2\rad/\clong) \phonaR_\phonindex^{-1} \kronecker_{\phonl\phonlb}\kronecker_{-\phonm,\phonmb}\kronecker_{\phonn\phonnb} \spare{1 - \phonm^2/(\phonl^2+\phonl)}/2\\
    $\phonfam = \Smode, \phonfamb=\Smode$ &
    \sqcoupling_{\phonindex\phonindexb} & = & (-1)^{\phonm+1} (\rotfreq^2\rad/\clong) \sqrt{\phonaR_\phonindex\phonaR_\phonindexb}^{-1} \kronecker_{-\phonm,\phonmb} \spare{ \kronecker_{\phonl\phonlb}\kronecker_{\phonn\phonnb}/2 + K_{-2} \,\kronecker_{\phonl-2,\phonlb}  + K_0 \,\kronecker_{\phonl\phonlb} + K_{+2}\, \kronecker_{\phonl+2,\phonlb} } \\
    $\phonfam = \Tmode, \phonfamb=\Smode$ &
    \sqcoupling_{\phonindex\phonindexb} & = & (-1)^{\phonm} (\rotfreq^2\rad/\clong) \sqrt{\phonaR_\phonindex\phonaR_\phonindexb}^{-1} \kronecker_{-\phonm,\phonmb} \spare{ K_{-1} \,\kronecker_{\phonl-1,\phonlb} + K_{+1} \,\kronecker_{\phonl+1,\phonlb}  }\\
    $\phonfam = \Smode, \phonfamb=\Tmode$ &
    \sqcoupling_{\phonindex\phonindexb} & = & \sqcoupling_{\phonindexb\phonindex}\\
    \bottomrule
  \end{tabularx}
  \setlength{\extrarowheight}{5pt}
  \newcolumntype{D}{>{\begin{math}\thinmuskip=.75mu\medmuskip=1mu}l<{\end{math}}}
  \begin{tabularx}{\textwidth}{ABD}
    K_{-2} & \equiv & -\spare{8 \phonl(\phonl-2) +6}^{-1} \sqrt{(2 \phonl-3) (\phonl-\phonm-1) (\phonl-\phonm) (\phonl+\phonm-1) (\phonl+\phonm)/(2 \phonl+1)}
    \Big\{ I_{\phonindex\phonindexb}^{\VSHYcomp\VSHYcomp} - (\phonl-2) I_{\phonindex\phonindexb}^{\VSHYcomp\VSHPsicomp} + (\phonl+1) I_{\phonindex\phonindexb}^{\VSHPsicomp\VSHYcomp} -(\phonl-2) (\phonl+1) I_{\phonindex\phonindexb}^{\VSHPsicomp\VSHPsicomp} \Big\} \\
    K_{-1} & \equiv & \im \phonm \sqrt{ (\phonl-\phonm) (\phonl+\phonm) / ( 4 \phonl^2-1) } \cpare{ I_{\phonindex\phonindexb}^{\VSHPhicomp\VSHYcomp} - (\phonl-1) I_{\phonindex\phonindexb}^{\VSHPhicomp\VSHPsicomp} } /2 \\
    K_{0} & \equiv & -[8 \phonl (\phonl+1)-6]^{-1} \Big\{ \spare{2 \phonl (\phonl + 1) - 2 \phonm^2 - 1} I_{\phonindex\phonindexb}^{\VSHYcomp\VSHYcomp} + \spare{\phonl (\phonl+1)-3 \phonm^2} \pare{I_{\phonindex\phonindexb}^{\VSHYcomp\VSHPsicomp} + I_{\phonindex\phonindexb}^{\VSHPsicomp\VSHYcomp}}  \spare{ 2 \phonl (\phonl+1) \pare{\phonl (\phonl+1)-\phonm^2}-3 \phonm^2} I_{\phonindex\phonindexb}^{\VSHPsicomp\VSHPsicomp}  \Big\}\\
    K_{+1} & \equiv & \im \phonm \sqrt{  (\phonl-\phonm+1) (\phonl+\phonm+1) / [ 4 \phonl(\phonl+2) +3] } \cpare{ I_{\phonindex\phonindexb}^{\VSHPhicomp\VSHYcomp} + (\phonl+2) I_{\phonindex\phonindexb}^{\VSHPhicomp\VSHPsicomp} } /2 \\
    K_{+2} & \equiv & -[4 \phonl+6]^{-1} \sqrt{ (\phonl-\phonm+1) (\phonl-\phonm+2) (\phonl+\phonm+1) (\phonl+\phonm+2) / [4 \phonl(\phonl+3) + 5] }
    \cpare{ I_{\phonindex\phonindexb}^{\VSHYcomp\VSHYcomp} + (\phonl+3) I_{\phonindex\phonindexb}^{\VSHYcomp\VSHPsicomp} - \phonl I_{\phonindex\phonindexb}^{\VSHPsicomp\VSHYcomp} - \phonl (\phonl+3) I_{\phonindex\phonindexb}^{\VSHPsicomp\VSHPsicomp} }  \\
    I_{\phonindex\phonindexb}^{ij} & \equiv & \int_0^1 \rposR^2 \wmodercomp^i_\phonindex(\rposR)\wmodercomp^j_\phonindexb(\rposR)\dd\rposR ~, \quad \text{for } i,j \in \{\VSHYcomp,\VSHPsicomp,\VSHPhicomp\} \\
    I_{\phonindex}^i & \equiv & \int_0^1 \rposR^3 \wmodercomp^i_\phonindex(\rposR) \dd \rposR~,  \quad \text{for } i \in \{ \VSHYcomp,\VSHPsicomp \}\\
    I^\VSHYcomp_\phonindex & = & A_\phonindex \spare{ \phonl I_4(\phonl,\phonaR_\phonindex) - \phonaR_\phonindex I_5(\phonl,\phonaR_\phonindex) } + C_\phonindex \phonl(\phonl+1) I_4(\phonbR_\phonindex)\\
    I^\VSHPsicomp_\phonindex & = & A_\phonindex I_4(\phonl,\phonaR_\phonindex) + C_\phonindex \spare{ (\phonl+1) I_4(\phonl,\phonbR_\phonindex) - \phonbR_\phonindex I_5(\phonl,\phonbR_\phonindex) }\\
    I_4(\phonl,\phonaR) & \equiv & \int_0^1 \rposR^2 \besselj{\phonl}(\phonaR \rposR) \dd \rposR = \sqrt{\pi} \phonaR^\phonl \spare{ 2^{\phonl+1} (\phonl+3) \Gamma\pare{\phonl+ 3/2 } }^{-1} ~ \generalizedHypergeometricF{1}{2}\spare{ (\phonl+3)/2; \, (\phonl+5)/2,\phonl+3/2; \, -\phonaR^2/4 } \\
    I_5(\phonl,\phonaR) & \equiv & \int_0^1 \rposR^3 \besselj{\phonl+1}(\phonaR \rposR) \dd \rposR = \sqrt{\pi} \phonaR^{\phonl+1} \spare{ 2^{\phonl+2} (\phonl+5) \Gamma\pare{\phonl+ 5/2 } }^{-1}  ~ \generalizedHypergeometricF{1}{2}\spare{ (\phonl+5)/2; \, (\phonl+7)/2,\phonl+5/2 ;\, -\phonaR^2/4 }\\
    \bottomrule
  \end{tabularx}
  \caption{Constants appearing in the Hamiltonian \cref{eqn: fixed rotation total Hamiltonian contributions 2} for a rotating sphere. The coupling constants have the symmetries $\sqcoupling_{\phonindexb\phonindex} = \sqcoupling_{\phonindex\phonindexb}$ and $\bscoupling_{\phonindexb\phonindex} = \cconj\bscoupling_{\phonindex\phonindexb}$. We distinguish coupling between different mode families $\phonfam$ and $\phonfamb$. The radial partial waves $\wmodercomp^i_\phonindex$ of the displacement modal fields of a sphere are given in \cref{tab: phonon displacement modal fields} in \cref{sec: sphere eigenmodes appendix} and the frequency-dependent radial constant $\phonaR_\phonindex$ and $\phonbR_\phonindex$ are defined in \cref{tab: phonon quantities}. The two integrals $I^\VSHYcomp_\phonindex$ and $I^\VSHPsicomp_\phonindex$ appearing in the linear shifts can be evaluated explicitly in terms of generalized hypergeometric functions $\tensor[_p]{F}{_q}$ and the gamma function $\Gamma$. The amplitudes $A_\phonindex$ and $C_\phonindex$ are specified in \cref{tab: phonon displacement modal field amplitudes}.}
  \label{tab: coupling constants sphere}
\end{table*}

\subsubsection{Spheroidal Modes}

Spheroidal modes are characterized by $B_\phonindex =0$. They are hybrid transverse and longitudinal excitations that cause displacement in all spatial directions and modify the nanoparticle shape. As in the case of the torsional modes, the boundary conditions \cref{eqn: boundary conditions matrix notation} lead to a discrete set of spheroidal modes for each $\phonl$ with eigenfrequencies degenerate in $\phonm$. We enumerate the eigenmodes using the radial index $\phonn$. Spheroidal modes can therefore be labeled by mode indices \mbox{$\phonindex = (\phonfam,\phonl,\phonm,\phonn)$} where
\begin{align}
  \phonfam&=\Smode, & \phonl &\in \N_0, & \phonm &\in \Z, |\phonm| \leq \phonl, & \phonn &\in \N
\end{align}
and we refer to each spheroidal eigenmode by a term of the form $\Smode_{\phonl\phonm\phonn}$. The spectrum of eigenfrequencies $\phonfreq_\phonindex$ derives from the roots $\phonbR_\phonindex  = (\clong/\ctrans) \phonaR_\phonindex$ of a transcendental frequency equation to be discussed. Unlike for torsional modes, the roots $\phonbR_\phonindex$ of the frequency equations of spheroidal modes are not universal but depend on the Poisson ratio $\Poissonnu$ only. We treat the two cases $\phonl=0$ and $\phonl \geq 1$ separately in order to obtain the respective frequency equation from the boundary conditions \cref{eqn: boundary conditions matrix notation}. In the case $\phonl=0$, the boundary conditions reduce to $A_\phonindex\, M_{11} = 0$. Hence, the frequency equation is
\begin{equation}
  4 \phonaR \besselj{1}(\phonaR) - \phonbR^2 \besselj{0}(\phonaR) = 0.
\end{equation}
In the case $\phonl \geq 1$, none of the coefficients $M_{ij}$ in \cref{tab: sphere phonon stress modal fields} vanish at all frequencies. In consequence, nontrivial solutions for the amplitudes $A_\phonindex$, $C_\phonindex=0$ can only exist if the determinant of the coefficient matrix of the first two equations in \cref{eqn: boundary conditions matrix notation} is zero. The frequency equation is hence
\begin{equation}\label{eqn: spheroidal mode frequency equation}
  M_{11}M_{23} - M_{21}M_{13} = 0
\end{equation}
in terms of the coefficients of the stress modal field defined in \cref{tab: sphere phonon stress modal fields}. We plot the spectrum of spheroidal modes in panel~(b) of \cref{fig: sphere phonon spectrum}.

The boundary conditions \cref{eqn: boundary conditions matrix notation} relate the two amplitudes $A_\phonindex$ and $C_\phonindex$, and the orthonormality condition \cref{eqn: phonon orthonormality condition} determines the norm of the remaining free amplitude. We choose the amplitudes to be real valued (see \cref{tab: phonon displacement modal field amplitudes}) such that the displacement modal field $\wmode_\phonindex$ of spheroidal modes also satisfies the symmetry \cref{eqn: displacement modal field azimuthal symmetry}. The displacement modal fields depend only on the Poisson ratio, the eigenfrequency, and the radius, and scale with the latter as $\rad^{-3/2}$, see \cref{tab: phonon displacement modal fields}.

In \cref{eqn: strain modal fields spheroidal modes}, we list the strain modal fields of spheroidal modes that we use in \cref{sec: case study} to predict changes in the optical properties of a spinning nanoparticle. Only the spheroidal $\Smode_{00\phonn}$ and $\Smode_{20\phonn}$ modes are affected by the static centrifugal force and contribute to the average strain $\averageStrain$. The spatial averages of their strain modal fields $\averageStrainMode_\phonindex$ are diagonal in Cartesian coordinates with diagonal elements
\begin{equation}\label{eqn: average strain S00n modes}
    \averageStrainModeComp_\phonindex^{ii} = \frac{A_\phonindex}{2 \sqrt{\pi \rad^5} } \frac{\phonaR_\phonindex  \cos (\phonaR_\phonindex )-\sin (\phonaR_\phonindex )}{ \phonaR_\phonindex  }
\end{equation}
for $i \in \{1, 2, 3\}$ in the case of the $\Smode_{00\phonn}$ modes and
\begin{multline}\label{eqn: average strain S2mn modes}
   \averageStrainModeComp_\phonindex^{ii} =
   \frac{A_\phonindex}{2 \sqrt{5 \pi  \rad^5} } \frac{\phonaR_\phonindex \cos (\phonaR_\phonindex ) - \sin (\phonaR_\phonindex )}{\phonaR_\phonindex }\\
   + \frac{3C_\phonindex}{2 \sqrt{5 \pi  \rad^5}} \frac{\phonbR_\phonindex \cos (\phonbR_\phonindex ) - \sin (\phonbR_\phonindex ) }{ \phonbR_\phonindex }
\end{multline}
for $i \in \{1, 2\}$ and $\averageStrainModeComp_\phonindex^{33} = -2 \averageStrainModeComp_\phonindex^{\xpos\xpos}$ in the case of the $\Smode_{20\phonn}$ modes.

\subsection{Spinning Sphere}
\label{sec: spinning sphere appendix}

The dynamics of a linear elastic sphere spinning at a constant frequency is governed by the Hamiltonian \cref{eqn: fixed rotation total Hamiltonian} with constants defined in \cref{eqn: coupling constants general}. We can obtain explicit expressions for these constants by using the displacement modal fields given in \cref{tab: phonon displacement modal fields}. The results are listed in \cref{tab: coupling constants sphere} and form the basis of the case study in \cref{sec: case study}.

The eigenmodes and eigenfrequencies of a spinning sphere can be constructed via a Bogoliubov transformation. For the sake of completeness, we provide a sketch of the procedure. It is useful to express the Hamiltonian in the matrix form
\begin{equation}
  \Hamilop = \frac{\hbar}{2}  \aopveccomp^i \Hamilmatrixcomp^{ij}  \aopveccomp^{j}
\end{equation}
which is equivalent to \cref{eqn: quadratic Hamiltonian} up to a constant. Here,
$\aopvec \equiv  \transp{(\displaced{\phonaop}_{1},\dots,\displaced{\phonaop}_{N},\hconj{\displaced{\phonaop}}_{1},\dots,\hconj{\displaced{\phonaop}}_{N})}$
is the $2N$-dimensional vector of ladder operators $\displaced{\phonaop}_\phonindex$ and we use $\phonindex$ both to denote the mode index $\phonindex=(\phonfam,\phonl,\phonm,\phonn)$ as well as to enumerate the $N$ lowest-frequency modes $\phonindex=1,\dots,N$. The Hamilton matrix $\Hamilmatrix \equiv \HamilmatrixDiagonal + \HamilmatrixHybridization $ is a complex-valued $(2N \times 2N)$-dimensional matrix with the two parts
\begin{align}
  \HamilmatrixDiagonal &\equiv \begin{bmatrix} \frequencyMatrix & \zerotens \\ \zerotens &  \frequencyMatrix \end{bmatrix}, &
  \HamilmatrixHybridization &\equiv \begin{bmatrix} \bscouplingMatrix & \cconj{\sqcouplingMatrix} \\ \sqcouplingMatrix &  \cconj{\bscouplingMatrix} \end{bmatrix}
\end{align}
where we define $(N \times N)$-dimensional submatrices $\frequencyMatrix$, $\bscouplingMatrix$, and $\sqcouplingMatrix$ with the eigenfrequencies $\frequencyMatrixComp_{\phonindex\phonindexb} \equiv \phonfreq_\phonindex \kronecker_{\phonindex\phonindexb}$ and the coupling constants $\bscoupling_{\phonindex\phonindexb}$ and $\sqcoupling_{\phonindex\phonindexb}$ as components, respectively. Direct diagonalization of the Hamilton matrix $\Hamilmatrix$ is not possible since this approach does not in general respect the bosonic structure~\cite{kustura_quadratic_2019}. Instead, we need to diagonalize the dynamical matrix $\dynamicalMatrix_\Hamilmatrix \equiv \JMatrix \Hamilmatrix$ where $\JMatrix \equiv \diagonal [ \id_N,  -\id_N ]$ and $\id_N$ is the $N$-dimensional identity matrix. The system is linearly stable if and only if the eigenfrequencies $\phonfreqInt_\phonindexInt$ are real-valued and occur in pairs $(\phonfreqInt_\phonindexInt, -\phonfreqInt_\phonindexInt)$ as eigenvalues of the dynamical matrix $\dynamicalMatrix_\Hamilmatrix$~\cite{kustura_quadratic_2019}. By calculating the eigenvalues of the dynamical matrix, we can therefore at once verify that the system is linearly stable and obtain the phonon spectrum $\phonfreqInt_\phonindexInt$ of the spinning nanoparticle.

Provided that the system is linearly stable the bosonic operators $\phonaopInt_\phonindexInt$ corresponding to the new eigenmodes can be constructed from the transformation matrix $\trafomatrix$ that diagonalizes the dynamical matrix. Let $\dynamicalMatrix_\HamilmatrixInt = \trafomatrix^{-1} \dynamicalMatrix_\Hamilmatrix \trafomatrix$ such that the transformed dynamical matrix is diagonal
$\dynamicalMatrix_\HamilmatrixInt =  \diagonal(\phonfreqInt_1,\dots,\phonfreqInt_N,-\phonfreqInt_1,\dots,-\phonfreqInt_N)$. The columns of the transformation matrix $\trafomatrix$ are formed by the unit eigenvectors $\unitvec(\phonfreqInt_\phonindexInt)$ and $\unitvec(-\phonfreqInt_\phonindexInt)$ of the dynamical matrix~\cite{kustura_quadratic_2019}: $\trafomatrixcomp_{ij} = \unitveccomp^i(\phonfreqInt_j)$ for $j \in [1,N]$ and $\trafomatrixcomp_{ij} = \unitveccomp^i(-\phonfreqInt_{j-N})$ for $j \in [N+1,2N]$. The eigenvectors of positive and negative eigenvalues are related. As a result, the transformation matrix obeys the symmetries $\trafomatrixcomp_{i(N+j)} = \cconj\trafomatrixcomp_{(N+i)j}$ and $\trafomatrixcomp_{(N+i)(N+j)} = \cconj\trafomatrixcomp_{ij}$ for $i,j \in [1,N]$. The ladder operators $\displaced{\phonaop}_\phonindex$ corresponding to eigenmodes of the resting nanoparticle are related to the ladder operators $\phonaopInt_\phonindexInt$ of the spinning nanoparticle through~\cite{kustura_quadratic_2019}
\begin{equation}
  \displaced{\phonaop}_\phonindex = \sum_{\phonindexInt} \spare{ \trafomatrixcomp_{\phonindex\phonindexInt} \phonaopInt_\phonindexInt +  \trafomatrixcomp_{\phonindex(N+\phonindexInt)} \hconj\phonaopInt_\phonindexInt }.
\end{equation}
In consequence, we can construct the displacement modal fields of the eigenmodes of the spinning nanoparticle as linear combinations of the modal fields of the resting nanoparticle
\begin{equation}
  \wmodeInt_\phonindexInt(\pos) = \sum_\phonindex \phonmodeoverlap_{\phonindexInt\phonindex} \wmode_\phonindex(\pos)
\end{equation}
where $\phonmodeoverlap_{\phonindexInt\phonindex} \equiv \sqrt{\phonfreqInt_\phonindexInt/\phonfreq_\phonindex} [\trafomatrixcomp_{\phonindex \phonindexInt} + \trafomatrixcomp_{(\phonindexcc+N)\phonindexInt } ]$. Here, $\phonindexcc$ is the mode index such that $\wmode_{\phonindexcc}(\pos) = \cconj{\wmode}_\phonindex(\pos)$. For a resting sphere,  taking the complex conjugate of the displacement modal fields given in \cref{tab: phonon displacement modal fields} corresponds to inverting the sign of the azimuthal order, $\phonindexcc = (\phonfam,\phonl,-\phonm,\phonn)$. The normalization of the new modal fields is then given by
\begin{equation}
  \int_\body \cconj{\wmodeInt}_\phonindexInt \cdot \wmodeInt_\phonindexIntb(\pos) \dd\pos = \sum_\phonindex \cconj\phonmodeoverlap_{\phonindexInt\phonindex} \phonmodeoverlap_{\phonindexIntb\phonindex} \simeq \kronecker_{\phonindexInt\phonindexIntb}
\end{equation}
compare \cref{eqn: phonon orthonormality condition}. The modal fields $\wmodeInt_\phonindexInt(\pos)$ are merely approximately orthonormal because we include only the $N$  lowest-frequency eigenmodes of the resting nanoparticle.

\section{Elastic Nonlinearity}
\label{sec: nonlinearity appendix}

\begin{figure}
  \begin{centering}
    \includegraphics[width=242.2156pt]{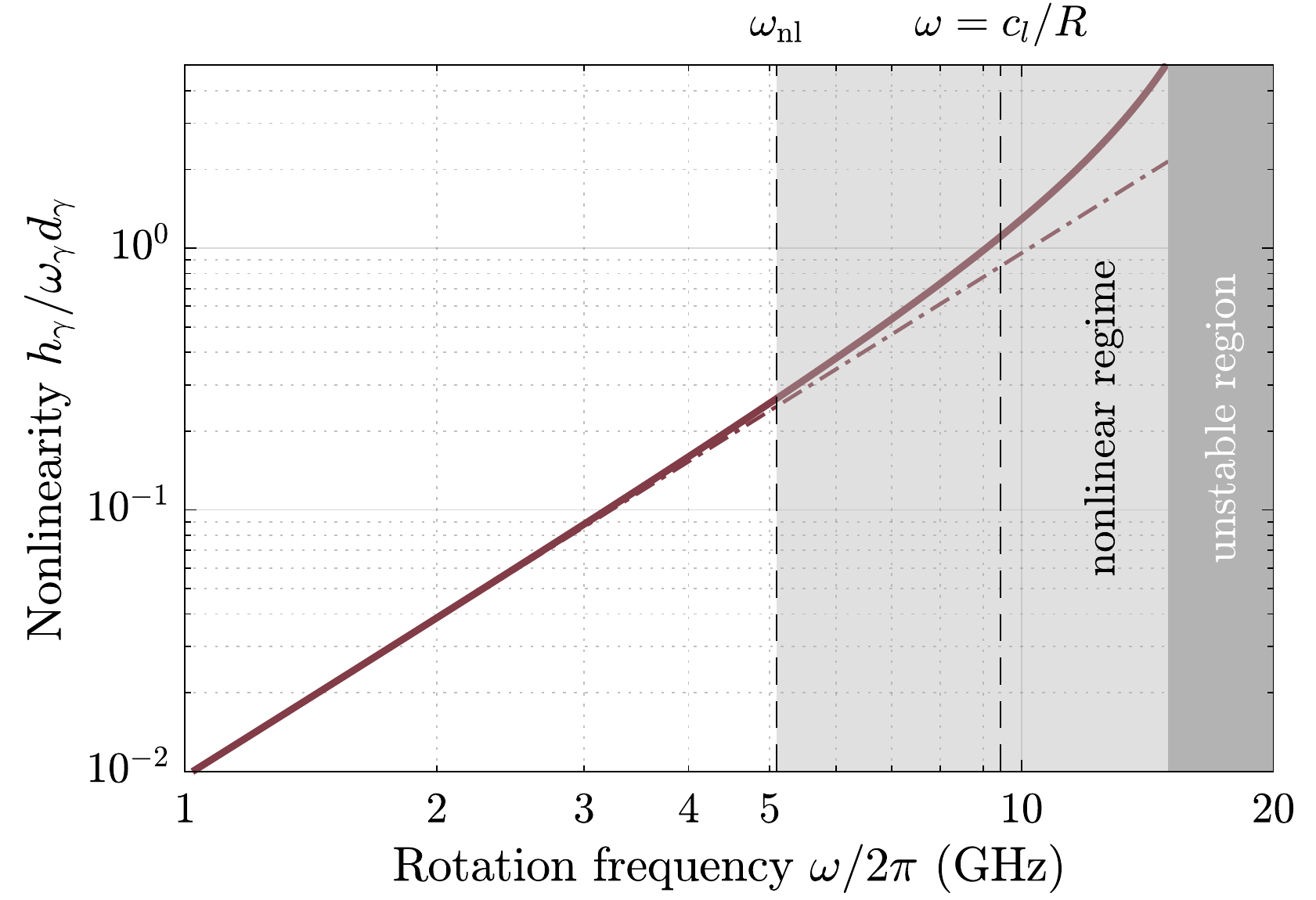}
  \end{centering}
  \caption{Relative contribution of anharmonic corrections to the elastic energy of the $\Smode_{001}$ mode of a spinning silica nanosphere, as defined in \cref{eqn: anharmonic corrections to elastic energy}. The dashed-dotted line indicates the power law $\aharmcorrection_\phonindex \phondispl_\phonindex / \phonfreq_\phonindex \propto \rotfreq^2 \rad^2 / \clong^2$ at frequencies $\rotfreq \ll \clong/\rad$.}
  \label{fig: mechanical nonlinearity}
\end{figure}

\begin{table*}
\setlength{\extrarowheight}{4pt}
  \newcolumntype{A}{>{\begin{math}}r<{\end{math}}}
  \newcolumntype{B}{>{\begin{math}}c<{\end{math}}}
  \newcolumntype{C}{>{\begin{math}}l<{\end{math}}}
  \begin{tabularx}{\textwidth}{ABCX}
    \toprule
    \elasenergydens_2(\nabla\ufield) & = & \lamemu \spare{ (\partial_i\ufieldcomp^{j}) (\partial_j\ufieldcomp^{i}) + (\del_i \ufieldcomp^{j}) (\del_j\ufieldcomp^{i}) }/2 + \lamelambda (\del_i\ufieldcomp^{i}) (\del_j \ufieldcomp^{j})/2 \\
    \elasenergydens_3(\nabla\ufield) & = & \anharmonicA  (\del_i\ufieldcomp^{j}) (\del_j\ufieldcomp^{k}) (\del_k\ufieldcomp^{i}) / 12  + (4\lamemu+\anharmonicA) (\del_i\ufieldcomp^{j}) (\del_j \ufieldcomp^{k}) (\del_i \ufieldcomp^{k}) /4 + \anharmonicB (\del_i\ufieldcomp^{i}) (\del_j\ufieldcomp^{k}) (\del_k\ufieldcomp^{j}) / 2 & ~ \\
    \multicolumn{4}{r}{$+ (\lamelambda+\anharmonicB) (\del_i\ufieldcomp^{i}) (\del_j\ufieldcomp^{k}) (\del_j\ufieldcomp^{k}) / 2 + \anharmonicC (\del_i\ufieldcomp^{i}) (\del_j \ufieldcomp^{j}) (\del_k\ufieldcomp^{k}) / 3$}\\
    \midrule
    \multicolumn{3}{l}{$\Smode_{001}$ mode}\\
    \aharmcorrection_\phonindex & = & \sqrt{2/\pi}  \sqrt{\hbar \clong/(\rad^6\dens)} \sqrt{\phonaR_\phonindex}^{-3} \left\{ 2 I(0,\phonaR_\phonindex) \spare{ \anharmonicAdimless + 6\anharmonicBdimless + 4\anharmonicCdimless + 3/(2-2\Poissonnu) }/3  +  I(1,\phonaR_\phonindex) \spare{2\anharmonicBdimless + 4\anharmonicCdimless + \Poissonnu/(1-\Poissonnu) } \right. \\
    \multicolumn{4}{r}{$+ \left. I(2,\phonaR_\phonindex) \spare{2\anharmonicBdimless + 2\anharmonicCdimless + \Poissonnu/(1-\Poissonnu) } + I(3,\phonaR_\phonindex) \spare{ 2\anharmonicAdimless + 6\anharmonicBdimless + 2\anharmonicCdimless + 3 } / 6  \right\}$}\\
    I(i,\phonaR) & \equiv & \int_0^1 \rposR^{i-1} \spare{\wmoder_\phonindex^{\VSHYcomp}(\rposR)}^{3-i} \spare{\partial_\rposR\wmoder_\phonindex^{\VSHYcomp}(\rposR)}^i \dd\rposR\\
    \bottomrule
  \end{tabularx}
  \caption{Elastic nonlinearity of a nanosphere. The terms $\elasenergydens_2(\nabla\ufield)$ and $\elasenergydens_3(\nabla\ufield)$ are the harmonic and leading-order anharmonic contributions to the elastic energy density $\elasenergydens$ of an isotropic body~\cite{landau_theory_1986}. The frequency $\aharmcorrection_\phonindex$ is the leading-order anharmonic correction to the frequency of the $\Smode_{001}$ mode. We define the dimensionless nonlinear elastic constants $\anharmonicAdimless \equiv \anharmonicA/(2\lamemu+\lamelambda)$ and likewise for $\anharmonicBdimless$ and $\anharmonicCdimless$.}
  \label{tab: elastic nonlinearity}
\end{table*}

The centrifugal strain experienced by a spinning nanoparticle increases with its rotation frequency. At extreme frequencies, anharmonic corrections to the interatomic interaction potential (elastic energy) $\potinternal$ become relevant. The resulting equation of motion of elasticity theory is then no longer linear, as opposed to \cref{eqn: phonon equation of motion}, and there can increasingly be deviations from the linear elastic theory presented in this paper. Here, we estimate at which rotation frequencies such elastic nonlinearities start to appear for a spinning nanosphere.

The deformation of a continuous body is described by its strain tensor $\straintens$. The full strain tensor has Cartesian components~\cite{landau_theory_1986}
\begin{equation}
  \straintenscomp^{ij} \equiv \frac{1}{2} \spare{  \del_i\ufieldcomp^{j} + \del_j\ufieldcomp^{i} + \sum_k \del_k\ufieldcomp^{i} \del_k\ufieldcomp^{j} }
\end{equation}
and includes terms quadratic in the displacement $\ufield$ that are neglected in linear elasticity theory; compare~\cref{eqn: definition strain and stress tensor}. The elastic energy density $\elasenergydens$ is a function of $\straintens$. For isotropic bodies, $\elasenergydens$ can only depend on the tensor invariants $\tensorinvar_1 = \tr (\straintens)$, $\tensorinvar_2 = \tr (\straintens^2)$, and $\tensorinvar_3 = \tr (\straintens^3)$~\cite{landau_theory_1986}. To third order in the strain
\begin{equation}
  \elasenergydens = \lamemu \tensorinvar_2 + \frac{\lamelambda}{2} \tensorinvar_1^2 + \frac{\anharmonicA}{3}\tensorinvar_3 + \anharmonicB \tensorinvar_1 \tensorinvar_2 + \frac{\anharmonicC}{3}\tensorinvar_1^3 + \order^4(\straintens).
\end{equation}
Here, $\lamemu$ and $\lamelambda$ are the linear elastic constants introduced in \cref{sec: case study} and $\anharmonicA$,  $\anharmonicB$, and $\anharmonicC$ are nonlinear elastic constants. We do not include a term constant in energy since it does not affect the dynamics. Moreover, there is no term linear in the strain, otherwise $\straintens=\zerotens$ would not be a minimum of the strain energy and $\ufield=0$ would not be an equilibrium configuration of the body. Keeping only terms up to third order in the displacement, we obtain~\cite{landau_theory_1986}
\begin{equation}
  \elasenergydens = \elasenergydens_2(\nabla\ufield) + \elasenergydens_3(\nabla\ufield) + \order^4(\nabla\ufield),
\end{equation}
where $\nabla\ufield$ is the Jacobi matrix of the displacement with Cartesian components $\nablacomp\ufieldcomp^{ij} = \del_i \ufieldcomp^j$. The harmonic term $\elasenergydens_2(\nabla\ufield)$ and the leading-order anharmonic correction $\elasenergydens_3(\nabla\ufield)$ are given in \cref{tab: elastic nonlinearity}.

The quantum operator representing the elastic energy of the body is hence $\elasenergyop = \elasenergyop_2 + \elasenergyop_3 + \order^4(\nabla\ufieldop)$, with
\begin{equation}
  \elasenergyop_i \equiv \int_B \elasenergydens_i(\nabla\ufieldop) \,\dd\pos
\end{equation}
for $i=2,3$. The linear elastic energy $\elasenergyop_2$ is contained in the Hamiltonian \cref{eqn: bare phonon Hamiltonian} of linear elasticity theory. The term $\elasenergyop_3$ represents the leading order nonlinear correction to the elastic energy. In general, it causes a shift in the eigenfrequencies of the linear elastic phonon eigenmodes summarized in \cref{sec: sphere appendix} and leads to coupling between the modes. The shifts and coupling frequencies can in principle be evaluated from $\elasenergydens_3$ in \cref{tab: elastic nonlinearity}. In practice, only the $\Smode_{001}$ and $\Smode_{201}$ are significantly displaced by the static centrifugal force and thereby primarily affected by anharmonicities; see \cref{sec: case study}. In order to obtain a simple criterion for the rotation frequencies at which anharmonic effects become relevant, we focus on the frequency shift of the radially symmetric $\Smode_{001}$ mode and neglect its anharmonic coupling to other modes. For the $\Smode_{001}$ mode
\begin{equation}
  \begin{split}
    \elasenergyop_2 &= \frac{\hbar \phonfreq_\phonindex}{4} (\phonaop_\phonindex^2 + 2\hconj{\phonaop}_\phonindex \phonaop_\phonindex + \phonaop_\phonindex^{\hcsymbol\,2}) \\
    \elasenergyop_3 &= \frac{\hbar \aharmcorrection_\phonindex}{8} (\phonaop_\phonindex^3 + 3 \hconj{\phonaop}_\phonindex \phonaop_\phonindex^2 + 3 \phonaop_\phonindex^{\hcsymbol\,2}\phonaop_\phonindex + \phonaop_\phonindex^{\hcsymbol\,3})
  \end{split}
\end{equation}
where we discard the energy of the vacuum state by normal ordering and define the anharmonic correction $\aharmcorrection_\phonindex$ to the eigenfrequency given in \cref{tab: elastic nonlinearity}. The expectation values with respect to the vacuum state $\ket{0}$ of the displaced mode $\displaced{\phonaop}_\phonindex$ (see \cref{sec: theory}) are $\braket{0|\elasenergyop_2|0} = \hbar \phonfreq_\phonindex \real^2(\phondispl_\phonindex)$ and $\braket{0|\elasenergyop_3|0} = \hbar \aharmcorrection_\phonindex \real^3(\phondispl_\phonindex)$. Here, the displacements $\phondispl_\phonindex\in \R$ such that we can express the expected elastic energy due to the $\Smode_{001}$ mode of a spinning nanoparticle in the absence of vibrations as
\begin{equation}\label{eqn: anharmonic corrections to elastic energy}
  \braket{0|\elasenergyop|0} = \hbar \phonfreq_\phonindex \phondispl_\phonindex^2 \spare{ 1 + \frac{\aharmcorrection_\phonindex}{\phonfreq_\phonindex}\phondispl_\phonindex + \order^2(\phondispl_\phonindex) }.
\end{equation}
Anharmonic effects can be neglected at rotation frequencies for which the frequency-dependent displacement $\phondispl_\phonindex$ is sufficiently small such that $\aharmcorrection_\phonindex \phondispl_\phonindex / \phonfreq_\phonindex \ll 1$. In \cref{fig: mechanical nonlinearity}, we plot $\aharmcorrection_\phonindex \phondispl_\phonindex / \phonfreq_\phonindex$ as a function of the rotation frequency for the parameters given in \cref{tab: parameters}. At frequencies $\rotfreq \ll \clong/\rad$, the anharmonic correction scales as $\aharmcorrection_\phonindex \phondispl_\phonindex / \phonfreq_\phonindex \propto \rotfreq^2 \rad^2 / \clong^2$ with a prefactor that only depends on the Poisson ratio and the three nonlinear elastic constants as indicated by the dashed-dotted line. We define $\rotfreqnl$ as the rotation frequency at which anharmonic corrections change the linear elastic energy of the displaced $\Smode_{001}$ by $\SI{25}{\percent}$. For the parameters specified in \cref{tab: parameters} and used in the case study in \cref{sec: case study}, we find that $\rotfreqnl = 2\times\SI{5.1}{\giga\hertz}$. We use $\rotfreqnl$ to indicate in which regime to expect sizable corrections to the linear elastic results presented in \cref{fig: change of shape,fig: permittivity and polarizability}.

\end{document}